\documentstyle[aps,multicol,eqsecnum,epsf,amsfonts]{revtex}

\begin{document}

\setlength{\unitlength}{1cm}
\def\B.#1{{\bbox#1}}
\newcommand{\Eq}[1]{Eq.~(\ref{#1})} \newcommand{\PM}{{\Bbb P}}
\newcommand{\PO}{\hat{\mathcal{P}}} \newcommand{\OO}{\hat{\mathcal{O}}}
\newcommand{\PRO}{\hat{\mathcal{P}}_{\mathrm{R}}}
\newcommand{\PLO}{\hat{\mathcal{P}}_{\mathrm{L}}} 
\newcommand{\KM}{{\Bbb K}} 
\newcommand{\KO}{\hat{\mathcal{K}}} 
\newcommand{\MM}{{\Bbb M}}
\newcommand{\OM}{{\Bbb O}} 
\newcommand{\LM}{{\Bbb L}}
\newcommand{\UM}{{\Bbb U}} 
\newcommand{\NM}{{\Bbb N}}
\newcommand{\Pjm}[1]{\Phi_{jm}(\B{#1})} 
\newcommand{\VecII}[2]{\left(\begin{array}{c} #1 \\ #2 \end{array} \right)}
\newcommand{\MatII}[4]{\left( \begin{array}{cc} #1 & #2 \\
      #3 & #4 \end{array}\right)}

\newcommand{\sepBR}[3]{\begin{picture}(9,1) \put(#2,#3){\line(1,0){#1}
      \line(0,1){.25}} \end{picture} }
\newcommand{\sepTL}[3]{\begin{picture}(9,1) \put(#2,#3){\line(0,-1){0.25}}
    \put(#2,#3){\line(1,0){#1}} \end{picture} }

\title{Spectrum of anisotropic exponents in hydrodynamic systems with
  pressure} 

\author{Itai Arad and Itamar Procaccia} 

\address{Deptartment of Chemical Physics, The Weizmann Institute of Science,
  Rehovot 76100, Israel} 

\date{version of December 29, 2000} 

\maketitle

%
%
\begin{abstract}  
  We discuss the scaling exponents characterizing the power-law behavior of
  the anisotropic components of correlation functions in turbulent systems
  with pressure. The anisotropic components are conveniently labeled by the
  angular momentum index $\ell$ of the irreducible representation of the
  SO(3) symmetry group. Such exponents govern the rate of decay of
  anisotropy with decreasing scales. It is a fundamental question whether
  they ever increase as $\ell$ increases, or they are bounded from above.
  The equations of motion in systems with pressure contain nonlocal
  integrals over all space. One could argue that the requirement of
  convergence of these integrals bounds the exponents from above. It is
  shown here on the basis of a solvable model (the ``linear pressure
  model"), that this is not necessarily the case. The model introduced here
  is of a passive vector advection by a rapidly varying velocity field. The
  advected vector field is divergent free and the equation contains a
  pressure term that maintains this condition. The zero modes of the
  second-order correlation function are found in all the sectors of the
  symmetry group. We show that the spectrum of scaling exponents can
  increase with $\ell$ without bounds, while preserving finite integrals.
  The conclusion is that contributions from higher and higher anisotropic
  sectors can disappear faster and faster upon decreasing the scales also
  in systems with pressure.
\end{abstract}

%
%

\begin{multicols}{2}
\section{Introduction}
Turbulent flows are often forced in an anisotropic fashion. The anisotropy
has a significant effect on a variety of measured turbulence
characteristics.  We are interested in the effect of anisotropy on
statistical quantities, especially the low-order structure functions of
velocity differences across a scale $R$. In perfectly isotropic systems
such objects are expected to display pure scaling behavior in the limit of
high Reynolds number. We have suggested recently \cite{99ALP} that in the
presence of anisotropy, the structure functions are no longer pure power
laws. Instead, components of the structure functions that belong to
different irreducible representations (sectors) of the SO($3$) group
possess different scaling exponents. Each of these sectors is characterized
by the angular momentum indices $\ell$ and $m$. By projecting the structure
function onto the different sectors, we could measure
\cite{98ADKLPS,99ABMP,00KLPS} the universal scaling exponents in each
sector separately.  

The spectrum of anisotropic exponents is particularly accessible in {\em
  linear} problems like passive scalar \cite{68Kra} and passively advected
magnetic fields \cite{96Ver}. In both the models, the equations of motion
are isotropic and as a result the existence of universal anisotropic
exponents can be proven \cite{00ALPP,00ABP}.  Additionally the isotropy of
the equations implies that the scaling exponents depend on $\ell$ but not
on $m$. One of the important results of the analysis is that the discrete
spectrum of anisotropic exponents is strictly increasing as a function of
$\ell$. This explains the isotropization of the statistics as smaller and
smaller scales are observed. Since the scaling exponents $\zeta$ appear in
power laws of the type $(R/L)^\zeta$, with $L$ being some typical outer
scale and $R \ll L$, the larger is the exponent, the faster is the decay of
the contribution as the scale $R$ diminishes. Therefore the gap between the
leading, isotropic exponent, and the next available exponent governs the
rate of isotropization.  

Experiments and simulations on Navier-Stokes (NS) turbulence also indicate
that anisotropic sectors possess larger scaling exponents than the
isotropic sector \cite{98ADKLPS,99ABMP,00KLPS}.  However, to date, the
exponents for $\ell>2$ were not determined with sufficient accuracy. We
thus do not know whether the higher sectors are characterized by an ever
increasing exponents, or whether the exponents {\em saturate}. This issue
is theoretically puzzling because of the effects of pressure. The inversion
of the pressure in terms of the Green's function of the Laplacian operator
introduces integrals over the domain of turbulence.  These integrals
manifest the nonlocality of the problem, and are present in both the
dynamical equation, and in the equations for the correlation functions.
When considering a spatially homogeneous turbulence, the turbulent domain
is usually taken to be infinite. The physical boundary conditions at scale
$L$, are mimicked by employing a homogeneous forcing that acts at that
scale. In this case, the integrals that result from the pressure term are
over all ${\Bbb R}^3$, and their convergence has to be guaranteed. The
question that we want to address in this paper is the following: does the
requirement of the convergence of the integrals necessarily bound the
spectrum of the scaling exponents from above? Since the correlation
functions appear in the integrand, an unbounded spectrum implies a rapidly
increasing integrand as a function of the length scale. On the face of it,
at some point the integrals must diverge in the infrared. It would appear
therefore that either there must be a limit to the magnitude of the scaling
exponents, or that the integrals converge due to an infrared crossover in
the correlation functions. The latter scenario looks physically reasonable,
yet in the presence of pressure integrals, seems to break scale invariance
in the inertial range. To demonstrate that, consider a typical integral
term of the form,
\begin{equation}
\label{eq:ex}
  \int\!\! d\B.y\, G(\B.r-\B.y) C(\B.y) \ .
\end{equation}
Here $G(\B.r) = -1/(4\pi|\B.r|)$ is the infinite domain Green function of
the Laplacian operator, and $C(\B.r)$ is some statistical object, which is
expected to be scale invariant in the inertial range. If $C(\B.r)$ has an
infrared crossover at scale $L$ (or equivalently, the integral has an
infrared cutoff at scale $L$), then the above expression will not be a pure
power law of $r$, not even inside the inertial range. Then how is it
possible that such an expression will cancel out a local term of $C(\B.r)$,
as is required by the typical equations of motion?

This puzzle has led in the past to the introduction of the concept of
``window of locality'' \cite{95LP,00FONGY}. The window of locality is the
range for the scaling exponents in which no divergence occurs, even if the
crossover length $L$ is taken to infinity. For these exponents, integrals
of type (\ref{eq:ex}) are dominated by the range of integration $y\approx
r$ and are therefore termed ``local". In a ``local" theory no infrared
cutoff is called for.

In this paper we show that scaling behavior of the correlation functions
together with finite integrals over an infinite domain {\em do not}
necessarily imply a bounded spectrum of anisotropic exponents. Our strategy
in this paper is to come up with a tractable example of a {\em linear}
model with pressure, see \Eq{eq:v-p}.  We refer to this model as the
``linear pressure model".  We approach the solutions of this model in two
steps. First we distill yet another, simpler, exactly solvable model, which
still poses the riddle of the Navier-Stokes problem. The exact solution
reveals that the spectrum of scaling exponents is unbounded, and the
convergence of the integrals is nevertheless not compromised. In the second
step we find the scaling exponents of the ``zero modes" \cite{95GK,95CFKL}
of the linear pressure model, and use the conclusions of the simpler model
to relate them to the full solution. We show that also in this case the
spectrum does not saturate in the anisotropic sectors.
                        
The linear pressure model and its simplified version reveal two mechanisms
that allow an unbounded spectrum of scaling exponents. First, a careful
analysis of the window of locality in the anisotropic sectors shows that it
widens as $\ell$ increases. We always have a leading scaling exponent
within the window of locality. Second, there is a more subtle mechanism
that comes to play when subleading exponents exist outside the window of
locality. In these cases we show that there exist counter-terms in the
exact solution (not the zero modes!) that maintain the locality of the
integrals. The bottom line is that in these models the anisotropic
exponents are unbounded from above leading to a fast decay of the
anisotropic contributions in the inertial range. 

In Sec.~\ref{sec:LinearP} we introduce the linear pressure model and derive
the equations for the two-point correlation function. We arrive at the form
containing the dangerous integrals, and discuss again the fundamental
riddle. In Sec.~\ref{sec:Toy} we construct a simpler, exactly solvable model
with the same riddle. In Sec.~\ref{sec:solution} we display the exact
solution of the model, and discuss the windows of locality and the
existence of an unbounded spectrum. In Sec.~\ref{sec:zeromodes} we go back
to the Linear Pressure model and offer a solution of its zero modes.
Section \ref{sec:summary} offers a summary and a discussion.

%
%

\section{Linear Pressure model}
\label{sec:LinearP}

%
%

\subsection{The model}
The linear pressure model captures some of the aspects of the pressure term
in Navier-Stokes turbulence, while being a linear and therefore, much
simpler problem. The nonlinearity of the Navier-Stokes equation is replaced
by an advecting field $\B.w(\B.x,t)$ and an advected field $\B.v(\B.x,t)$.
The advecting field $\B.w(\B.x, t)$ is taken with known dynamics and
statistics.  Both the fields are assumed incompressible. The equation of
motion for the vector field $v^\alpha(\B.x, t)$ is:
\begin{eqnarray}
    \partial_t v^\alpha + w^\mu \partial_\mu v^\alpha +
        \partial^\alpha p - \kappa \partial^2 v^\alpha &=& f^\alpha \ ,
        \label{eq:v-p} \\
    \partial_\alpha v^\alpha &=& 0 \ , \\
    \partial_\alpha w^\alpha &=& 0 \ .
\end{eqnarray}
In this equation, $\B.f(\B.x,t)$ is a divergence free forcing term and
$\kappa$ is the viscosity. The domain of the system is taken to be
infinite.  Following Kraichnan's model for passive scalar \cite{68Kra}, we
choose the advecting field $\B.w(\B.x,t)$ to be a white-noise Gaussian
process with a correlation function that is given by
\begin{eqnarray}
    \delta(t'-t)D^{\alpha\beta}(\B.r) &\equiv& 
        \left< w^\alpha(\B.x + \B.r, t') w^\beta(\B.x, t) \right> \ ,\\
     K^{\alpha\beta}(\B.r) &\equiv& 
         D^{\alpha\beta}(\B.r) - D^{\alpha\beta}(\B.0) \\
      &=& Dr^\xi\left[(\xi+2)\delta^{\alpha\beta} - 
               \xi\frac{r^\alpha r^\beta}{r^2}\right] \ .
\end{eqnarray}
The forcing $\B.f(\B.x,t)$ is also taken to be a Gaussian white noise
process. Its correlation function is
\begin{equation}
  F^{\alpha\beta} (\B.r/L)\delta(t-t') \equiv 
   \langle f^\alpha(\B.x+\B.r,t) f^\beta(\B.x,t')\rangle \ .
\end{equation}
The forcing is responsible for injecting energy and anisotropy to the
system at an outer scale $L$. We choose the tensor function
$F^{\alpha\beta}(\B.x)$ to be analytic in $\B.x$, anisotropic, and
vanishing rapidly for $|\B.x|\gg 1$. Analyticity is an important
requirement. It means that $F^{\alpha\beta}(\B.x)$ can be expanded for
small $|\B.x|$ as a power series in $x^\alpha$; as a result its leading
contribution in the $\ell$ sector is proportional to $x^{\ell-2}$, given by
$\partial^\alpha\partial^\beta x^\ell Y_{\ell m}(\hat{\B.x})$. To see that
this is the leading contribution the reader can consult the general
discussion of the construction of the irreducible representations in
Ref.\cite{99ALP}. All other analytic contributions contain less derivatives
and are therefore of higher order in $x$.  

`In order to derive the statistical equations of the correlation function of
$v^\alpha(\B.x,t)$, we need a version of \Eq{eq:v-p} without the
pressure term. Following the standard treatment of the pressure term in
Navier-Stokes equation, we take the divergence of \Eq{eq:v-p} and arrive
at
\begin{equation}
    \partial_\nu \partial_\mu w^\mu v^\nu +
        \partial^2 p = 0 \ .
\end{equation}
The Laplace equation is now inverted using the Green function of infinite
domain with zero-at-infinity boundary conditions,
\begin{eqnarray}
  p(\B.x) &=& -\int\!\! d\B.y\, G (\B.x - \B.y) 
     \partial_\nu \partial_\mu w^\mu(\B.y) v^\nu(\B.y) \ , \\
  G(\B.x) &\equiv& -\frac{1}{4\pi|\B.x|} \ . \label{eq:green}
\end{eqnarray}
With this expression for $p(\B.x)$, Eq.~(\ref{eq:v-p}) can be rewritten as
\begin{eqnarray}
\label{eq:v-no-p}
  && \partial_t v^\alpha(\B.x, t) 
     + w^\mu(\B.x,t) \partial_\mu v^\alpha(\B.x,t) \\
  && \quad - \partial^\alpha_{(\B.x)} \int\!\! d\B.y\, 
    G(\B.x-\B.y)\partial_\nu\partial_\mu w^\mu(\B.y) v^\nu(\B.y) \\
   && \quad - \kappa \partial^2 v^\alpha(\B.x,t) = f^\alpha(\B.x,t) \ . 
\nonumber
\end{eqnarray}

%
%

\subsection{Equations for the second-order correlation function}
\label{sec:C}
The statistical object that we are interested in is the two-point
correlation function of the field $v^\alpha(\B.x)$,
\begin{equation}
    C^{\alpha\beta}(\B.r) \equiv 
       \left<v^\alpha(\B.x+\B.r) v^\beta(\B.x) \right> \ .
\end{equation}
We find its equation of motion in two steps. First, we take the time
derivative of $C^{\alpha\beta}(\B.r)$ using \Eq{eq:v-no-p},
\begin{eqnarray}
  && \partial_t\left<v^\alpha(\B.x+\B.r)v^\beta(\B.x)\right> + 
  \left<v^\alpha(\B.x+\B.r)w^\mu(\B.x)\partial_\mu v^\beta(\B.x)\right> \\ 
  && + \left<v^\beta(\B.x)w^\mu(\B.x+\B.r)\partial_\mu 
       v^\alpha(\B.x+\B.r)\right> \\
  && - \left<v^\alpha(\B.x+\B.r)\partial^\beta_{(x)}
         \int\!\!d\B.y\, G(\B.x-\B.y)\partial_\mu\partial_\nu 
         w^\mu(\B.y) v^\nu(\B.y) \right> \nonumber \\
  && - \left<v^\beta(\B.x)\partial^\alpha_{(x+r)}
         \int\!\!d\B.y \, G(\B.x+\B.r-\B.y)\partial_\mu\partial_\nu 
         w^\mu(\B.y) v^\nu(\B.y) \right> \nonumber \\
  && - \kappa\left<v^\alpha(\B.x+\B.r)\partial^2v^\beta(\B.x)\right> 
     - \kappa\left<v^\beta(\B.x)\partial^2v^\alpha(\B.x+\B.r)\right>
     \nonumber \\
  && = \left<v^\alpha(\B.x+\B.r) f^\beta(\B.x)\right> + 
       \left<v^\beta(\B.x) f^\alpha(\B.x+\B.r)\right> \ . \nonumber
\end{eqnarray}
To simplify the equations we define an auxiliary function
$T^{\alpha\beta}(\B.r)$
\begin{equation}
    \label{eq:T-definition} 
  T^{\alpha\beta}(\B.r) \equiv
    \partial_\mu^{(r)} \left< v^\alpha(\B.x+\B.r) w^\mu(\B.x) v^\beta(\B.x)
      \right> \ . 
\end{equation}
Using this definition and the space homogeneity of the statistics, we
arrive after some algebraic manipulation to the following equation:
\begin{eqnarray}
  && \partial_t C^{\alpha\beta}(\B.r) - T^{\alpha\beta}(\B.r)
    -T^{\beta\alpha}(-\B.r) \\
  && + \int\!\!d\B.y\, G(\B.r - \B.y) \partial^\beta\partial_\nu 
     T^{\alpha\nu}(\B.y)
  + \int\!\!d\B.y\, G(-\B.r - \B.y) \partial^\alpha\partial_\nu
        T^{\beta\nu}(\B.y) \nonumber \\
  && -2\kappa\partial^2 C^{\alpha\beta}(\B.r) = 
   \left<v^\alpha(\B.x+\B.r) f^\beta(\B.x)\right> + 
       \left<v^\beta(\B.x) f^\alpha(\B.x+\B.r)\right> \ . \nonumber
\end{eqnarray}

The last equation is identical to the equation for the second-order
correlation function in the usual Navier-Stokes turbulence, provided that
$w^\mu$ is replaced with $v^\mu$ in \Eq{eq:T-definition}. Indeed, the
vexing problem that we face is being made very clear: if the triple
correlation function has a power law dependence on $\B.r$ with an
arbitrarily large exponent, how can the integral converge in the infrared?
One possibility is that the scaling exponent of $T^{\alpha\beta}(\B.r)$ is
sufficiently low, making the integral convergent. The other possibility is
that the correlation function is scale invariant only in the inertial range
and vanishes quickly after that, which is equivalent to the introduction of
an infrared cutoff. However, the integral terms in the equation probe the
correlation function throughout the entire space. Therefore, a crossover
behavior of the correlation function at the outer scale $L$, seems to
contradict a pure scaling behavior of the correlation function in the
inertial range itself. This in turn implies the saturation of the
anisotropic scaling exponents.  

To proceed, we use the fact that the field $\B.w(\B.x,t)$, as well as the
forcing, are Gaussian white noises. This enables us to express
$T^{\alpha\beta}(\B.r)$ and the correlation of the force in terms of
$C^{\alpha\beta}(\B.r)$ and $F^{\alpha\beta}(\B.r)$. In appendix
\ref{sec:gauss} we use the well-known method of Gaussian integration by
parts \cite{ref:Frisch} that leads to the final equations,
\begin{eqnarray}
  \partial_t C^{\alpha\beta}(\B.r) &=& T^{\alpha\beta}(\B.r)
    + T^{\beta\alpha}(-\B.r) \label{eq:dtC} \\
   &&- \int\!\!d\B.y\, G(\B.r - \B.y) \partial^\beta\partial_\nu 
      T^{\alpha\nu}(\B.y) \nonumber \\
   &&- \int\!\!d\B.y\, G(-\B.r - \B.y) \partial^\alpha\partial_\nu
         T^{\beta\nu}(\B.y)  \nonumber \\
   &&+ 2\kappa\partial^2 C^{\alpha\beta}(\B.r) + 
     F^{\alpha\beta}(\B.r) \ , \nonumber  \\
  T^{\alpha\beta}(\B.r) &=& -\frac{1}{2}K^{\mu\nu}\partial_\mu\partial_\nu
     C^{\alpha\beta}(\B.r) \label{eq:Tab} \\
  &&+ \frac{1}{2}\partial^\alpha_{(\B.r)} \int\!\! d\B.y\, G(\B.r-\B.y)
     \partial_\tau\Big[ K^{\mu\nu}(\B.y) \partial_\mu \partial_\nu 
      C^{\tau\beta}(\B.y) \Big] \nonumber  \\
  &&- \frac{1}{2} \int \!\! d\B.y\, G(\B.y) \partial^\beta\partial_\tau 
    \Big[ K^{\mu\nu}(\B.y)\partial_\mu\partial_\nu
        C^{\alpha\tau}(\B.r-\B.y)\Big] \ . \nonumber 
\end{eqnarray}
These equations have to be supplemented with two more equations that follow
directly from the definition of $C^{\alpha\beta}(\B.r)$,
\begin{eqnarray}
\label{eq:C-Incomp} 
  \partial_\alpha C^{\alpha\beta}(\B.r) &=& 0 \ , \\
   C^{\alpha\beta}(\B.r) &=& C^{\beta\alpha}(-\B.r) \ . 
\label{eq:C-symmetry}
\end{eqnarray}
The first equation follows from the incompressibility constraint of the
vector field $\B.v(\B.x,t)$, while the latter follows from space
homogeneity. Finally, we note that Eqs.~(\ref{eq:dtC}) and (\ref{eq:Tab})
can be interpreted in a transparent way, utilizing two projection operators
that maintain the right-hand side (RHS) of \Eq{eq:dtC} divergence-free in
both indices.  To define them, let us consider a tensor field
$X^{\alpha\beta}(\B.r)$ that vanishes sufficiently fast at infinity. Then
the two projection operators $\PLO$ and $\PRO$ are defined by
\begin{eqnarray}
  \PLO X^{\alpha\beta}(\B.r) &\equiv& X^{\alpha\beta}(\B.r) - 
     \partial^\alpha_{(r)}\int d\B.y\, G(\B.r-\B.y)
        \partial_\mu  X^{\mu\beta}(\B.y) \ , \\
  \PRO X^{\alpha\beta}(\B.r) &\equiv& X^{\alpha\beta}(\B.r) - 
     \partial^\beta_{(r)}\int d\B.y\, G(\B.r-\B.y)
         \partial_\mu  X^{\alpha\mu}(\B.y) \ .
\end{eqnarray}
We observe that $\PLO X^{\alpha\beta}$ and $\PRO X^{\alpha\beta}$ are
divergence-free in the left and right indices, respectively. Using these
operators we can rewrite Eqs.(\ref{eq:dtC}-\ref{eq:Tab}) in the form
\begin{eqnarray}
  \partial_t C^{\alpha\beta}(\B.r) &=& 
    \PRO T^{\alpha\beta}(\B.r) + \PRO T^{\beta\alpha}(-\B.r)
    \label{eq:dtC1} \\
   &+& 2 \kappa\partial^2 C^{\alpha\beta}(\B.r)
    + F^{\alpha\beta}(\B.r) \ , \nonumber \\
  T^{\alpha\beta}(\B.r) &=& -\frac{1}{2} \PLO K^{\mu\nu}\partial_\mu\partial_\nu
    C^{\alpha\beta}(\B.r) - \label{eq:Tab1} \\ 
    &-& \frac{1}{2} \int \!\! d\B.y\, G(\B.y) \partial^\beta\partial_\tau 
    \Big[ K^{\mu\nu}(\B.y)\partial_\mu\partial_\nu
        C^{\alpha\tau}(\B.r-\B.y)\Big] \ . \nonumber
\end{eqnarray}
The projection in \Eq{eq:Tab1} guarantees that $T^{\alpha\beta}(\B.r)$ is
divergence free in its left index, while the projection in \Eq{eq:dtC1}
guarantees divergence freedom in the right index.

Not all the terms in these equations are of the same nature. The integrals
due to the projection operator are easy to deal with by applying a
Laplacian on them. For example,
\begin{equation}
  \partial^2 \PRO T^{\alpha\beta}(\B.r) = \partial^2 T^{\alpha\beta}(\B.r)
     - \partial^\beta \partial_\nu T^{\alpha\nu}(\B.r) \ .
\end{equation}
On the other hand, there seems to be no way to eliminate the last integral
in \Eq{eq:Tab1}, and therefore we shall refer to it as the ``nontrivial
integral''. Only for $\xi=0$ and $\xi=2$ it trivializes: the integral
vanishes when $\xi=0$ and is proportional to $C^{\alpha\beta}(\B.r)$ when
$\xi=2$.  Unfortunately, in these extreme cases also the projection
operator trivializes, and the effect of the pressure cannot be adequately
assessed. We prefer to study the problem for a generic value $\xi$ for
which the incompressibility constraint and the pressure terms are
nontrivial.

We deal with the this problem head-on in Sec.~\ref{sec:zeromodes}. Due to
the nontrivial integral, we will not be able to provide a full solution of
$C^{\alpha\beta}(\B.r)$, but only of the zero modes.  However, before doing
so we would like to study a model that affords an exact solution in order
to understand in detail the issues at hand. In the next section we
therefore consider a simplified model of the linear pressure model, yet
posing much of the same riddle.

%
%

\section{An exactly solvable toy model}
\label{sec:Toy}
We construct a toy model that inspired by Eqs.~ (\ref{eq:dtC}) and,
(\ref{eq:Tab}) for the correlation function in the linear pressure model.
Within this model we demonstrate the strategy of dealing with the
nonlocal-pressure term. Since it is a simplification of the {\em
  statistical} equation of the linear pressure model, the toy model has no
obvious underlying dynamical equation.

%
%

\subsection{Definition of the toy model}

In the toy model, we are looking for a ``correlation function''
$C^\alpha(\B.r)$, whose equations of motion are
\begin{eqnarray}
  \partial_t C^{\alpha}(\B.r) &=& 
     -K^{\mu\nu}(\B.r)\partial_\mu\partial_\nu C^\alpha(\B.r) -
     \partial^\alpha_{(r)} \label{eq:C} \\
 && \times \int \!\! d\B.x G(\B.r-\B.x) \partial_\tau 
      K^{\mu\nu}(\B.x)\partial_\mu\partial_\nu C^\tau(\B.x)  \nonumber \\ 
 && + \ \kappa \partial^2 C^\alpha(\B.r)+F^\alpha(\B.r/L)  \ , \nonumber \\
\partial_\alpha C^\alpha(\B.r) &=& 0 \label{eq:toy-incomp} \ .
\end{eqnarray}
Here $F^\alpha(\B.x)$ is a one-index analog of the correlation function of
the original forces $F^{\alpha\beta}(\B.x)$. Accordingly, we take it
anisotropic, analytic in $x^\alpha$ and rapidly vanishing for $|\B.x| \gg
1$. As in the previous model, also here analyticity requires that the
leading contribution for small $|\B.x|$ is proportional to $\partial^\alpha
x^\ell Y_{\ell m}(\hat{\B.x})$ in the $\ell$ sector. Accordingly it is of
order $x^{\ell-1}$. 

The toy model is simpler than the Linear Pressure model in two aspects:
First, the ``correlation function'', $C^\alpha(\B.r)$ has one index instead
of two and therefore can be represented by a smaller number of scalar
functions. Second, the unpleasant nontrivial term of the linear pressure
model is absent. This will allow us to solve the model exactly for every
value of $\xi$. Nevertheless, the toy model confronts us with the same
conceptual problems that exist in the linear pressure model and in NS
turbulence: can a scale invariant solution in the inertial range with a
crossover to a decaying solution at scale $L$, be consistent with the
integral term? If not, is there a saturation of the anisotropic exponents?

Equation \ref{eq:C} can be rewritten in terms of a new projection operator
$\PO$, which projects a vector $X^\alpha(\B.r)$ on its divergence free
part
\begin{equation}
  \partial_t C^{\alpha} = 
    -\PO \Big[ K^{\mu\nu}\partial_\mu\partial_\nu C^\alpha \Big] 
     + \kappa \partial^2 C^\alpha +F^\alpha  \ ,
\label{eq:C-with-P}
\end{equation}
where
\begin{equation}
  \PO X^\alpha(\B.r) \equiv X^\alpha(\B.r) - 
    \partial^\alpha \int\!\! d\B.y\, G(\B.r-\B.y)\partial_\mu X^\mu(\B.y) \ .
\label{def:P}
\end{equation}

We shall solve this integro-differential equation by first turning it into
a partial differentail equation (PDE) using the Laplacian operator, and
then turning it into a set of decoupled ODE's using the SO($3$)
decomposition. 

As in the linear pressure model, the nonlocality of the projection operator
can be removed by considering a differential version of it
\begin{equation}
  \partial^2 \PO T^\alpha(\B.r) = \partial^2 T^\alpha(\B.r) - 
     \partial^\alpha \partial_\mu T^\mu(\B.r) \ .
\end{equation}
In stationary condition $\partial_t C^\alpha=0$, and therefore the
differential form of the toy model is given by
\begin{eqnarray}
 && \partial^2 \PO 
  \Big[K^{\mu\nu}(\B.r)\partial_\mu\partial_\nu C^\alpha(\B.r)\Big] \nonumber \\
 && \quad = \partial^2 K^{\mu\nu}(\B.r)
      \partial_\mu\partial_\nu C^\alpha(\B.r) - \partial^\alpha \partial_\tau 
         K^{\mu\nu}(\B.r)\partial_\mu\partial_\nu C^\tau(\B.r) \nonumber \\
 && \quad = \kappa \partial^2\partial^2 C^\alpha +\partial^2 F^\alpha  \ ,
         \label{eq:diff-C} \\
 && \partial_\alpha C^\alpha(\B.r) = 0 \ . 
\end{eqnarray}
We have reached a linear PDE of order four. This PDE will be solved by
exploiting its symmetries, i.e., isotropy and parity conservation, as
demonstrated in the next subsection.

%
%

\subsection{The SO($3$) decomposition}

Equation \ref{eq:diff-C} and the incompressibility condition of
$C^\alpha(\B.r)$ are both isotropic and parity conserving. Therefore, if we
expand $C^\alpha(\B.r)$ in terms of spherical vectors with a definite
behavior under rotations and under reflections, we would get a set of
decoupled ODE's for their coefficients.  For each sector $(\ell,m),\ 
\ell>0$ of SO($3$) we have three spherical vectors:
\begin{eqnarray}
  A_1^\alpha(\hat{\B.r}) &\equiv& r^{-\ell-1}r^\alpha \Phi_{\ell m}(\B.r) \ ,
    \nonumber \\
  A_2^\alpha(\hat{\B.r}) &\equiv& 
      r^{-\ell+1}\partial^\alpha \Phi_{\ell m}(\B.r) \ ,
    \nonumber \\
  A_3^\alpha(\hat{\B.r}) &\equiv& r^{-\ell} 
    \epsilon^{\alpha\mu\nu}r_\mu\partial_\nu \Phi_{\ell m}(\B.r) \ .
\label{def:A1-3}
\end{eqnarray}
Here $\Phi_{\ell m}(\B.r)=r^\ell Y_{\ell m}(\hat{\B.r})$, and see
\cite{99ALP} for further details. The first two spherical vectors have a
different parity than the third vector, hence the equations for their
coefficients are decoupled from the equation for the third coefficient. In
the following, we shall consider the equations for the first two
coefficients only, as they have a richer structure and larger resemblance
to the linear pressure model. Finally, note that the isotropic sector, i.e.,
$\ell=0$, is identically zero. To see why, notice that in this special
sector there is only one spherical vector, $A_1^\alpha(\hat{\B.r}) \equiv
r^{-1}r^\alpha$.  Hence the isotropic part of $C^\alpha(\B.r)$ is given by
$c(r)r^{-1}r^\alpha$, $c(r)$ being some scalar function of $r$. But then
the incompressibility condition (\ref{eq:toy-incomp}) implies that $c(r)
\sim r^{-2}$, which has a UV divergence. We therefore conclude that
$c(r)=0$, and restrict our calculation to $\ell>0$.  

By expanding $C^\alpha(\B.r)$ in terms of the spherical vectors $\B.A_1$
and $\B.A_2$, we obtain a set of ODEs [decoupled in the $(\ell,m)$ labels]
for the scalar functions that are the coefficients of these vectors in the
expansion. The equations for these coefficients can thus be written in
terms of matrices and column vectors. To simplify the calculations, we find
the matrix forms of the Kraichnan operator and of the Laplacian of the
projection operator separately, and only then combine the two results to
one.

%
%
\subsubsection{The matrix form of the Kraichnan operator}
To obtain the matrix of the Kraichnan operator in the basis of $\B.A_1$ and
$\B.A_2$, we expand $C^\alpha(\B.r)$:
\begin{equation}
  C^\alpha(\B.r) = 
    c_1(r)A_1^\alpha(\hat{\B.r}) + c_2(r)A_2^\alpha(\hat{\B.r}) \ .
\end{equation}
Using the basic identities of the $\Phi_{\ell m}(\B.r)$ functions (see
\cite{99ALP}),
\begin{eqnarray*}
  \partial^2 \Phi_{\ell m}(\B.r) &=& 0 \ , \\
  r^\mu \partial_\mu \Phi_{\ell m}(\B.r) &=& \ell \Phi_{\ell m}(\B.r) \ ,
\end{eqnarray*}
a short calculation yields
\begin{eqnarray}
   && \KO C^\alpha(\B.r) \equiv
    K^{\mu\nu}(\B.r)\partial_\mu\partial_\nu C^\alpha(\B.r) \\
   && \quad = Dx^\xi\Big[ 2c''_1 + 2(2+\xi)\frac{c'_1}{r} - 
                (2+\xi)(\ell+1)(\ell+2)\frac{c_1}{r^2}
              \Big]A_1^\alpha(\hat{\B.r}) \nonumber \\
     && \quad \ + Dx^\xi\Big[ 2c''_2 + 2(2+\xi)\frac{c'_2}{r}
        + 2(2+\xi)\frac{c_1}{r^2}  \nonumber \\
     && \quad \ - 2(2+\xi)\ell(\ell-1)\frac{c_2}{r^2} \Big]
     A_2^\alpha(\hat{\B.r}) \nonumber \ .
\end{eqnarray}

Therefore, in matrix notation, the Kraichnan operator can be written as:
\begin{eqnarray}
 && \KO \VecII{c_1}{c_2} = 2Dr^\xi \MatII{1}{0}{0}{1}\VecII{c''_1}{c''_2} 
     \nonumber \\
 && \quad  + 2D(2+\xi)r^{\xi-1} \MatII{1}{0}{0}{1}\VecII{c'_1}{c'_2} 
     \nonumber \\
 && \quad - D(2+\xi)r^{\xi-2} \MatII{(\ell+1)(\ell+2)}{0}{-2}{\ell(\ell-1)}
      \VecII{c_1}{c_2} \nonumber \\
 && \equiv r^\xi \KM_2\VecII{c''_1}{c''_2} + 
          r^{\xi-1}\KM_1\VecII{c'_1}{c'_2} + 
          r^{\xi-2}\KM_0\VecII{c_1}{c_2} \ .
\label{def:K-mat}
\end{eqnarray}
%
%
\subsubsection{The matrix form of the Laplacian of the Projection operator}
Let
\begin{equation}
  T^\alpha(\B.r) = t_1(r)A_1^\alpha(\hat{\B.r}) + 
    t_2(r)A_2^\alpha(\hat{\B.r}) 
\end{equation}
and applying a Laplacian to $\PO T^\alpha$, we get
\begin{eqnarray}
\label{eq:explicit-P}
  && \partial^2\PO T^\alpha =
    \Big[ -\ell t''_2 + \ell\frac{t'_1}{r} + \ell(2\ell-1)\frac{t'_2}{r}\\
  && \quad  - \ell(\ell+1)\frac{t_1}{r^2} 
     - \ell(\ell-1)(\ell+1)\frac{t_2}{r^2}\Big]A_1^\alpha \nonumber \\ 
  && + \left[ t''_2 - \frac{t'_1}{r} + (2-\ell)\frac{t'_2}{r}\right]
     A_2^\alpha \ . \nonumber
\end{eqnarray}
Hence in matrix notation,
\begin{eqnarray}
 && \partial^2\PO\VecII{t_1}{t_2} = \MatII{0}{-\ell}{0}{1}
        \VecII{t''_1}{t''_2} \nonumber\\
   && \quad  + \frac{1}{r}\MatII{\ell}{\ell(2\ell-1)}{-1}{2-\ell}
      \VecII{t'_1}{t'_2} \nonumber \\
   && \quad - \frac{1}{r^2}\MatII{\ell(\ell+1)}{\ell(\ell-1)(\ell+1)}{0}{0}
      \VecII{t_1}{t_2}   \nonumber \\
   && \quad \equiv \PM_2\VecII{t^{''}_1}{t^{''}_2} + \frac{1}{r}\PM_1
      \VecII{t'_1}{t'_2} + \frac{1}{r^2}\PM_0\VecII{t_1}{t_2} \ .
\label{def:P-mat}
\end{eqnarray}

%
%

\subsection{The matrix form of the toy model}

Now that the matrix forms of the Kraichnan operator and of the Laplacian of
the projection operator have been found, we can combine these two results
to find the matrix form of the left-hand side (LHS) of \Eq{eq:diff-C}. To
this aim let us define
\begin{equation}
  \VecII{t_1}{t_2} = \KO \VecII{c_1}{c_2} \ 
\end{equation}
and from Eqs.~(\ref{def:K-mat}) and (\ref{def:P-mat}) we get
\begin{eqnarray}
  && \partial^2 \PO \KO \VecII{c_1}{c_2} =
    r^\xi \MM_4 \VecII{c^{(4)}_1}{c^{(4)}_2} + 
    r^{\xi-1} \MM_3 \VecII{c^{(3)}_1}{c^{(3)}_2} \\
  && \quad +\ r^{\xi-2} \MM_2 \VecII{c^{(2)}_1}{c^{(2)}_2} 
   + r^{\xi-3} \MM_1 \VecII{c^{(1)}_1}{c^{(1)}_2} \nonumber \\ 
&&\quad +\ r^{\xi-4} \MM_0 \VecII{c_1}{c_2} \ , \nonumber
\end{eqnarray}
where the number in parentheses denotes the order of the derivative. The
matrices $\MM_i$ are given by
\begin{eqnarray}
\label{def:M-mat}
\MM_4 &\equiv& \PM_2\KM_2 \ , \\
\MM_3 &\equiv& 2\xi\PM_2\KM_2 + \PM_2\KM_1 + \PM_1\KM_2 \ , \nonumber \\
\MM_2 &\equiv& \xi(\xi-1)\PM_2\KM_2 + 2(\xi-1)\PM_2\KM_1 + \PM_2\KM_0 
         \nonumber \\
      && + \xi\PM_1\KM_2 + \PM_1\KM_1 + \PM_0\KM_2 \ , \nonumber \\
\MM_1 &\equiv& (\xi-1)(\xi-2)\PM_2\KM_1 + 2(\xi-2)\PM_2\KM_0 \nonumber \\
      && + (\xi -1)\PM_1\KM_1 +\PM_1\KM_0 + \PM_0\KM_1 \ ,\nonumber \\
\MM_0 &\equiv& (\xi-2)(\xi-3)\PM_2\KM_0 + (\xi-2)\PM_1\KM_0 
          + \PM_0\KM_0  \ . \nonumber
\end{eqnarray}

To find the RHS of \Eq{eq:diff-C} we expand the ``forcing''
$F^\alpha(\B.r)$ in terms of the spherical vectors $\B.A_1$ and $\B.A_2$,
\begin{equation}
  F^\alpha(\B.r) = f_1(r) A^\alpha_1(\hat{\B.r}) + 
                  f_2(r) A^\alpha_2(\hat{\B.r}) \ ,
\end{equation}
and applying a Laplacian we find the matrix form of
$\partial^2 F^\alpha(\B.r)$,
\begin{eqnarray}
&& \partial^2 \VecII{f_1}{f_2} = 
   \left( \begin{array}{c}
    f''_1 + \frac{2}{r}f'_1 - (\ell+1)(\ell+2)\frac{1}{r^2}f_1 \\ \ \\
   f''_2 + \frac{2}{r}f'_2 + \frac{2}{r^2}f_1 - \ell(\ell-1)\frac{1}{r^2}f_2
   \end{array} \right) \nonumber \\
&&\quad \equiv \VecII{\rho_1}{\rho_2} \ .
\end{eqnarray}
At this point it is worthwhile to remember that the forcing term
$F^\alpha(\B.r/L)$ is assumed to be analytic. As a result for $r/L \ll 1$,
its leading contribution in the $(\ell,m)$ sector is proportional to
$\partial^\alpha r^\ell Y_{\ell m}(\hat{\B.r}) \sim r^{\ell-1}$. However
$\partial^2 F^{\alpha}(\B.r/L)$ is also analytic, and must therefore also
scale like $r^{\ell-1}$ for small $r$, instead of scaling like
$r^{\ell-3}$, which could be the naive dimensional guess. 

To proceed we restrict ourselves to finding the solution in the inertial
range and beyond. In these ranges the dissipative term $\kappa
\partial^2\partial^2 C^\alpha(\B.r)$ is negligible and can be omitted, thus
reaching the following equation for the column vector $(c_1, c_2)$:
\begin{eqnarray}
  && r^\xi \MM_4 \VecII{c^{(4)}_1}{c^{(4)}_2} + 
    r^{\xi-1} \MM_3 \VecII{c^{(3)}_1}{c^{(3)}_2} + 
    r^{\xi-2} \MM_2 \VecII{c^{(2)}_1}{c^{(2)}_2}  \nonumber \\
  && \quad + r^{\xi-3} \MM_1 \VecII{c^{(1)}_1}{c^{(1)}_2} + 
    r^{\xi-4} \MM_0 \VecII{c_1}{c_2} =   \VecII{\rho_1}{\rho_2} \ .
\label{eq:C-mat}
\end{eqnarray}

Finally, also the incompressibility constraint $\partial_\alpha
C^\alpha(\B.r)=0$, can be expressed as a relation between $c_1(r)$ and
$c_2(r)$,
\begin{equation}
\label{eq:incomp}
  c'_1 + 2\frac{c_1}{r} + \ell c'_2 - \ell(\ell-1)\frac{c_2}{r} = 0 \ .
\end{equation}
This constraint has to be taken into account when solving \Eq{eq:C-mat}.

%
%

\section{Solving the toy model}
\label{sec:solution}

%
%

\subsection{The general solution}

The solution of \Eq{eq:C-mat} is somewhat tricky due to the additional
constraint (\ref{eq:incomp}). Seemingly the two unknowns $c_1(r)$ and $
c_2(r)$ are over determined by the three equations (\ref{eq:C-mat},
\ref{eq:incomp}), yet this is not the case for the two equations
(\ref{eq:C-mat}) are not independent (when considered as two scalar
equations, resulting from the two-column vectorial equation). To see that
this is the case and find the solution, it is advantageous to work in the
new basis
\begin{eqnarray}
  d_1 &=& c_1 + \ell c_2 \ , \nonumber \\
  d_2 &=& -2c_1 + \ell(\ell-1)c_2 \ .
\label{eq:def-d}
\end{eqnarray}
In this basis the incompressibility constraint becomes very simple,
\begin{equation}
  d_2 = rd_1' 
\end{equation}
allowing us to express $d_2$ and its derivatives in terms of $d_1$. To do
that in the framework of the matrix notation, we define the transformation
matrix $\UM$
\begin{eqnarray}
  \UM &\equiv& \MatII{1}{\ell}{-2}{\ell(\ell-1)} \ , 
\label{def:U-mat} \\
  \UM^{-1} &=& \frac{1}{\ell(\ell+1)}\MatII{\ell(\ell-1)}{-\ell}{2}{1} \ ,
\label{def:invU-mat} 
\end{eqnarray}
so that,
\begin{equation}
   \VecII{d_1}{d_2} = \UM \VecII{c_1}{c_2} \ .
\end{equation}
The equations of $d_i(r)$ are the same as the equations for $c_i(r)$, with
the matrices $\MM_i$ replaced by
\begin{equation}
  \NM_i \equiv \UM \MM_i \UM^{-1}
\label{def:N-mat}
\end{equation}
and the sources $\rho_i$ replaced by
\begin{equation}
  \VecII{\rho^*_1}{\rho^*_2} = \UM \VecII{\rho_1}{\rho_2} \ .
\end{equation}
Notice that a divergence free forcing $F^\alpha(\B.r)$ will cause
$\rho^*_1(r)$ and $\rho^*_2(r)$ to be related to each other in the same way
that $d_1(r)$ and $d_2(r)$ are related to each other, i.e.,
\begin{equation}
  \rho^*_2 = r(\rho^*_1)' \ .
\end{equation}
Next, we perform the following replacements:
\begin{eqnarray*}
  d_2       &=& rd^{(1)}_1 \ , \\
  d_2^{(1)} &=& rd^{(2)}_1 + d^{(1)}_1 \ , \\
  d_2^{(2)} &=& rd^{(3)}_1 + 2d^{(2)}_1 \ , \\
  d_2^{(3)} &=& rd^{(4)}_1 + 3d^{(3)}_1 \ , \\
  d_2^{(4)} &=& rd^{(5)}_1 + 4d^{(3)}_1 \ .
\end{eqnarray*}
We get an equation written entirely in terms of the function $d_1(r)$ and
its derivatives,
\begin{eqnarray}
  &&r^{\xi+1} V_5 d^{(5)}_1 + r^\xi V_4 d^{(4)}_1 + r^{\xi-1}V_3 d^{(3)}_1 
  + r^{\xi-2}V_2 d^{(2)}_1 \nonumber\\
 && + r^{\xi-3}V_1 d^{(1)}_1 + r^{\xi-4}V_0 d_1 =
    \VecII{\rho^*_1}{\rho^*_2} \ , 
\label{eq:V5}
\end{eqnarray}
where $V_i$ are two dimensional vectors given by,
\begin{eqnarray}
  V_5 &\equiv& \NM_4 \VecII{0}{1} \ , \label{def:V} \\
  V_4 &\equiv& \NM_4 \VecII{1}{4} + \NM_3\VecII{0}{1} \ , \nonumber \\
  V_3 &\equiv& \NM_3 \VecII{1}{3} + \NM_2\VecII{0}{1} \ , \nonumber \\
  V_2 &\equiv& \NM_2 \VecII{1}{2} + \NM_1\VecII{0}{1} \ , \nonumber \\
  V_1 &\equiv& \NM_1 \VecII{1}{1} + \NM_0\VecII{0}{1} \ , \nonumber \\
  V_0 &\equiv& \NM_0 \VecII{1}{0}  \ . \nonumber \\
\end{eqnarray}
\end{multicols}
\sepBR{8.6}{-0.3}{.5} \\
Their explicit values are given by,
\begin{eqnarray*}
  V_5 &=& D\VecII{0}{2} \ , \\
  V_4 &=& D\VecII{2}{16+6\xi} \ , \\
  V_3 &=& D\VecII{16+4\xi}{-4\ell^2 - 4\ell + 32 \xi 
    + 8 - \xi \ell^2 - \xi \ell + 6\xi^2} \ , \\
  V_2 &=& D\VecII{-4\ell^2 - 4\ell + 20\xi + 24 - \xi\ell^2 -\xi\ell + 2\xi^2}
                {-8\xi\ell^2 - 8\xi\ell - 4\xi - 48 + 22\xi^2 
                  -2\xi^2\ell^2 - 2\xi^2\ell + 2\xi^3} \ , \\
  V_1 &=& D\VecII{-(\xi +2)(-6\xi + \xi\ell^2 + \xi\ell + 4\ell^2 + 4\ell)}
                {(\xi+2)(6\xi^2 - \xi^2\ell^2 - \xi^2\ell- \xi\ell^2 - 18\xi 
                 - \xi\ell + \ell^4 + 11\ell^2 + 2\ell^3 + 10\ell)} \ , \\
  V_0 &=& D\VecII{\ell(\ell-1)(\ell+2)(\ell+1)(\xi+2)}
                {\ell(\xi+2)(\xi-4)(\ell-1)(\ell+2)(\ell+1)} \ .
\end{eqnarray*}
\rightline{\sepTL{8.7}{0.4}{0}}
\begin{multicols}{2}
Equation \ref{eq:V5} is for a column vector, and can be regarded as two
scalar differential equations that we refer to as the ``upper" and the
``lower". The upper ODE is of the fourth order, while the lower ODE is of
fifth order. Not surprisingly, the lower equation is the first derivative
of the upper equation, provided that $F^\alpha(\B.r)$ is divergence free.
Hence the two equations are dependent, and we restrict our attention to the
upper equation.  To simplify it, we divide both sides by $Dr^\xi$, replace
$d_1(r)$ by $\psi(r)$ and define the RHS to be the function $S(r)$,
\begin{equation}
  S(r)\equiv D^{-1}r^{-\xi}\rho^*_1(r) \ .
\label{eq:S(r)}
\end{equation}
After doing so, we reach the following equation:
\begin{equation}
  \psi^{(4)} + a_3\frac{\psi^{(3)}}{r} + a_2\frac{\psi^{(2)}}{r^2} +
    a_1 \frac{\psi^{(1)}}{r^3} + a_0 \frac{\psi}{r^4} =S(r) \ .
\label{eq:psi}
\end{equation}
Its homogeneous solution is easily found once we substitute,
\begin{equation}
  \psi(r) = \psi_0 r^\zeta \ .
\end{equation}
The scaling exponents are the roots of the polynomial,
\begin{eqnarray}
\label{def:Polynomial}
   P(\zeta) &=& \zeta(\zeta-1)(\zeta-2)(\zeta-3)  \\
   &+& a_3\zeta(\zeta-1)(\zeta-2) + a_2\zeta(\zeta-1) + a_1\zeta + a_0 
\nonumber
\end{eqnarray}
and are found to be real and nondegenerate. Two of them are positive while
the other two are negative, given in a decreasing order by
\begin{eqnarray}
\label{def:exponents}
  \zeta_1 &=& -\frac{1}{2} - \frac{1}{2}\xi + \frac{1}{2}
     \sqrt{A(\ell,\xi) + \sqrt{B(\ell,\xi)}} \ , \\
  \zeta_2 &=& -\frac{1}{2} - \frac{1}{2}\xi + \frac{1}{2}
     \sqrt{A(\ell,\xi) - \sqrt{B(\ell,\xi)}} \ , \nonumber \\
  \zeta_3 &=& -\frac{1}{2} - \frac{1}{2}\xi - \frac{1}{2}
     \sqrt{A(\ell,\xi) - \sqrt{B(\ell,\xi)}} \ , \nonumber \\
  \zeta_4 &=& -\frac{1}{2} - \frac{1}{2}\xi - \frac{1}{2}
     \sqrt{A(\ell,\xi) + \sqrt{B(\ell,\xi)}} \ , \nonumber
\end{eqnarray}
where
\begin{eqnarray}
  A(\ell,\xi) &\equiv& \xi^2 +\xi \ell^2 + \xi \ell - 2\xi + 5 + 4\ell 
    + 4\ell^2 \ , \\
  B(\ell,\xi) &\equiv&  -8\xi^2\ell - 7\xi^2\ell^2 + 16\xi^2 + 2\xi^2\ell^3
    + \xi^2\ell^4 \\
    && - 8\xi \ell^2 - 8\xi \ell - 32\xi + 16 + 64\ell +64\ell^2 \ .
\nonumber
\end{eqnarray}
In the limit $\xi \to 0$ the roots become:
\begin{eqnarray}
\label{def:zero-xi-exponents}
   \zeta_1 &=& \ell+1  \ , \\
   \zeta_2 &=& \ell-1  \ , \nonumber \\
   \zeta_3 &=& -\ell   \ , \nonumber \\
   \zeta_4 &=& -\ell-2 \ . \nonumber
\end{eqnarray}
Figure 1 displays the first few exponents as a function of $\xi$.  We note
that the spectrum has no sign of saturation as $\ell$ increases. Before we
discuss the meaning of this observation we will make sure that these
solutions are physically relevant and participate in the full (exact)
solution including boundary conditions. 

The general solution of \Eq{eq:psi} is traditionally given as the sum of a
special solution of the non-homogeneous equation plus a linear combination
of the zero modes. However, when attempting to match the solution to the
boundary conditions, it is more convenient to represent it as
\begin{equation}
\label{eq:psi-general-solution}
 \psi(r) = \sum_{i=1}^4 
  \frac{r^{\zeta_i}}{\underbrace{ (\zeta_i-\zeta_1) \cdots 
       (\zeta_i-\zeta_4)}_{\scriptsize\mbox{all {\em different} roots}}}
   \int_{m_i}^r \!\! dx \, x^{3-\zeta_i}\, S(x) \ ,
\end{equation}
where the free parameters of the solution are the four constants $m_i$.
Indeed a change in $m_i$ is equivalent to adding to the solution a
const$\times r^{\zeta_i}$. In the next subsection we find the values of
$m_i$ that match the boundary conditions, and discuss the properties of the
solution.
\end{multicols}
\begin{figure}[h]
\label{fig:exponents}
\epsfysize=10.5cm
\epsfbox{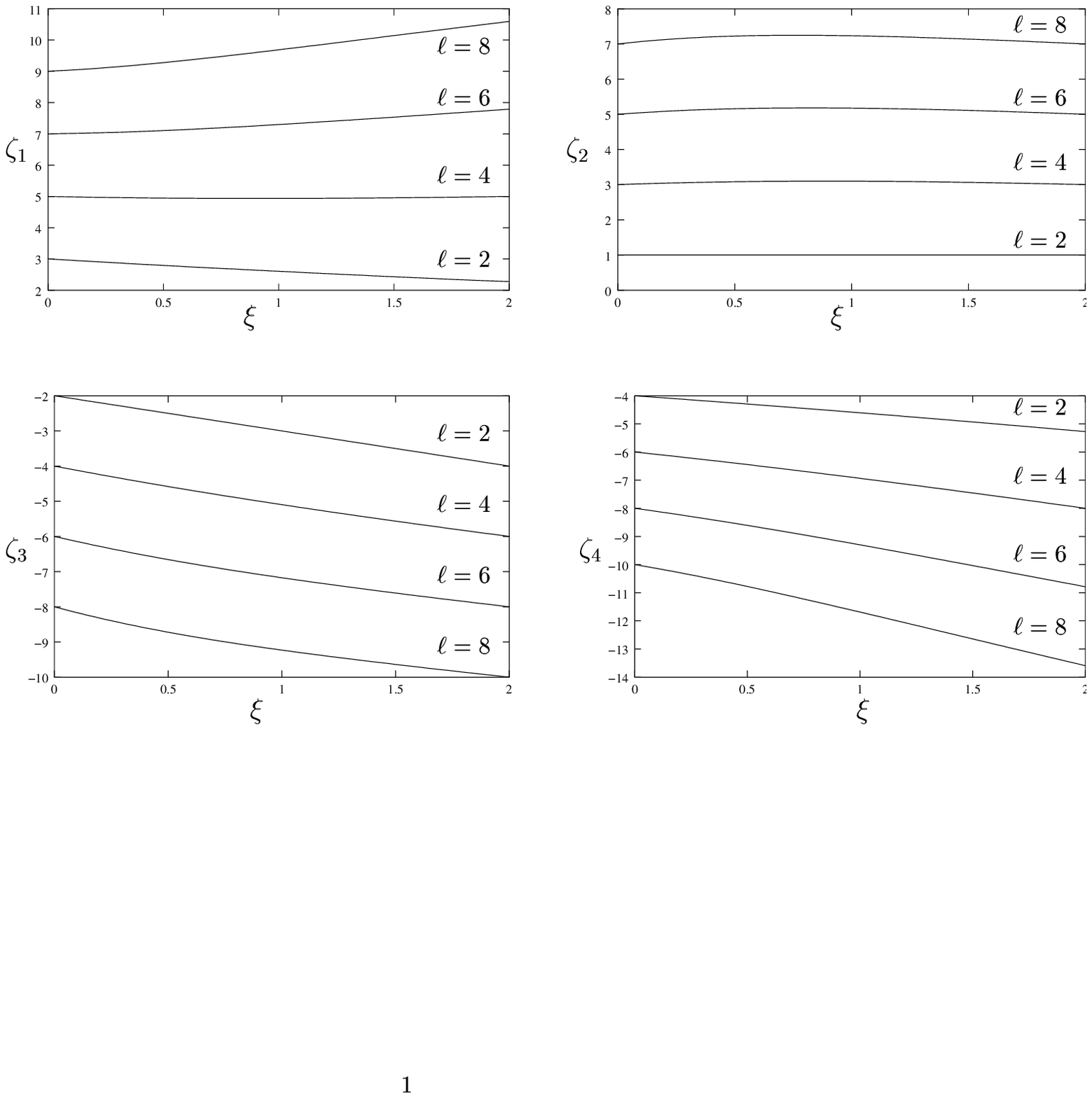}
\caption{Scaling exponents of the first few $\ell$'s as functions of $\xi$}
\end{figure}

\begin{multicols}{2}

%
%

\subsection{Boundary conditions and inertial-range behavior}
\label{sec:toy-bc}

From \Eq{eq:psi-general-solution} it is clear that the only values of $m_i$
that guarantee that the solution remains finite as $r\to 0$ and that it
decays as $r\to\infty$ are $m_1=m_2=+\infty$, $m_3=m_4 = 0$
\begin{eqnarray}
&\psi&(r) = 
  -\frac{r^{\zeta_1}}{(\zeta_1-\zeta_2)(\zeta_1-\zeta_3)(\zeta_1-\zeta_4)} 
  \int_r^\infty \!\! dx \, x^{3-\zeta_1}\, S(x) \nonumber\\
&-&\frac{r^{\zeta_2}}{(\zeta_2-\zeta_1)(\zeta_2-\zeta_3)(\zeta_2-\zeta_4)} 
  \int_r^\infty \!\! dx \, x^{3-\zeta_2 }\, S(x) \nonumber\\
&+&\frac{r^{\zeta_3}}{(\zeta_3-\zeta_1)(\zeta_3-\zeta_2)(\zeta_3-\zeta_4)} 
  \int_0^r \!\! dx \, x^{3-\zeta_3 }\, S(x) \nonumber\\
&+&\frac{r^{\zeta_4}}{(\zeta_4-\zeta_1)(\zeta_4-\zeta_2)(\zeta_4-\zeta_3)} 
  \int_0^r \!\! dx \, x^{3-\zeta_4}\, S(x) \ . 
\label{eq:fullsol}
\end{eqnarray}
To understand the asymptotics of this solution we find from \Eq{eq:S(r)}
that for $x \ll L$, $S(x)$ has a leading term that goes like
$x^{\ell-1-\xi}$, whereas for for $x \gg L$, $S(x)$ decays rapidly. It is
now straightforward to prove that for $r \ll L$, the $\zeta_3$ and
$\zeta_4$ terms scale like $r^{\ell+3-\xi}$, the $\zeta_2$ term scales like
$r^{\zeta_2}$, and the $\zeta_1$ term scales like $r^{\zeta_1}$ for values
of $\xi$ for which $\zeta_1 < \ell+3-\xi$ and like $r^{\ell+3-\xi}$
otherwise. In addition, it is easy to see that for $r \gg L$, $\psi(r)$
exhibits an algebraic decay: the $\zeta_1$ and $\zeta_2$ terms decay
rapidly due to the decay of $S(x)$ whereas the $\zeta_3$ and $\zeta_4$
terms decay algebraicaly like $r^{\zeta_i}$, respectively. The asymptotics
of the full solution are thus given by
\begin{equation}
  \psi(r) \sim \left\{ \begin{array}{lcr}
    r^{\zeta_2}, && r \ll L \\
    r^{\zeta_3}, && r \gg L \end{array} \right. \ .
\end{equation}

The obvious conclusion is that there is no saturation in the anisotropic
scaling exponents as $\ell$ increases. The lack of contradiction with the
existence of an integral over all space has two aspects. The main one is
simple and obvious. The integro-differential equation (\ref{eq:C}) for
$C^\alpha$ has a differential version (\ref{eq:diff-C}). Solving the
differential version, we are unaffected by any considerations of convergence
of integrals and therefore the solution may contain exponents that increase
with $\ell$ without limit. Nevertheless, the full solution
(\ref{eq:fullsol}) exhibits a crossover at $L$: it increases in the
inertial range $r\ll L$ and decays for $r\gg L$. Thus plugging it back to
the integro-differential equation we are guaranteed that no divergence
occurs. 

The question why the cross-over length $L$ does not spoil the scale
invariance in the intertial range still remains. The answer is found in
differential form of the equation of motion, given by \Eq{eq:diff-C}.  From
this equation we find that the integrand is a Green's function times a
Laplacian of a tensor. By definition, such an integral localizes, i.e., it
is fully determined by the value of the tensor at the external vector
$\B.r$.  In the language of \Eq{eq:ex} $A(\B.y)=\nabla^2 B(\B.y)$

The second and less obvious aspect is that the window of locality widens up
with $\ell$. This is due to the cancellations in the angular integration of
the anisotropic solutions that are due to the orthogonality of the $Y_{\ell
  m}(\hat{\B.r})$ and their generalizations $B^{\alpha}_{q \ell
  m}(\hat{\B.r})$. To demonstrate this consider again the simple integral
(\ref{eq:ex}), and assume that $C(\B.y)$ belongs to $(\ell, m)$ sector,
i.e.,
\begin{equation}
  C(\B.y) =a(y) Y_{\ell m}(\hat{\B.y}) \ .
\end{equation}
For $y \gg r$, we may expand the Green function in $r/y$,
\begin{eqnarray}
  G(\B.r-\B.y) &=& -\frac{1}{4\pi|\B.r-\B.y|} \\
   &=& -\frac{1}{4\pi y}\sum_{n=0}^\infty
     a_n \left[\left(\frac{r}{y}\right)^2 
       - 2\frac{\B.r \cdot \hat\B.y}{y}\right]^n \ . \nonumber
\end{eqnarray}
Here $a_n$ are Taylor coefficients. Obviously the dangerous terms for the
infrared convergence are those with low values of $n$. However, all these
terms will vanish for $n<\ell$ in the angular integration against
$Y_{\ell m}(\hat{\B.y})$. The reason is that all these terms are of the
form $r^{n_1} y^{n_2} (\B.r\cdot \hat\B.y)^{n_3}$ with $n_3< \ell$. The
angular part here has projections only $Y_{\ell' m'}$ with $\ell'\le
k_3<\ell$.  The first term to contribute comes when $n=\ell$, and is
proportional to the amplitude integral $\int_r^\infty \!\! dy \, y^2
a_{\ell m}(y) y^{-\ell -1}$.  For a power law $a_{\ell m}(y)\sim y^\lambda$
this implies locality for
\begin{equation}
  \lambda< \ell-2
\end{equation}
instead of $\lambda<-2$, as in the isotropic sector. The lower bound of the
window of locality is also extended and a similar analysis for $y\ll r$
leads to $\lambda>-\ell-3$. For the toy model this translates to the window
of locality
\begin{equation}
  -\ell-\xi < \zeta_i < \ell +1 -\xi \ .
\end{equation}
From the previous analysis we find that the leading power law of the full
solution in the inertial range is $r^{\zeta_2}$, which is inside this
``extended'' window of locality. Nevertheless, the subleading power
$r^{\zeta_1}$ originating from the first term in \Eq{eq:fullsol} is above
this window, and its presence in the solution can be explained only using
the first mechanism. 

We will see when we turn back to the linear pressure model that both these
mechanisms operate there as well, leading again to a lack of saturation in
the exponents.

%
%

\section{Solving the Linear Pressure model}
\label{sec:zeromodes}

We now return to the linear pressure model. As mentioned before in
Sec.~\ref{sec:C}, we see no way of eliminating all integrals from the
equation and therefore we will not look for a full solution. Nevertheless,
we shall be able to calculate the zero modes and hence the scaling
exponents. Our strategy relies heavily on the conclusions of the last
section: we will apply two Laplacians to the equation for
$C^{\alpha\beta}(\B.r)$ in order to eliminate the integrals of the two
projection operators $\PLO$ and $\PRO$. The resulting equation will still
contain the nontrivial integral. Using numerical integration we will solve
the homogeneous part of this equation, i.e., we shall find its zero modes.
These are scale invariant solutions that solve an equation containing an
integral. Their exponent must therefore lie within the ``extended'' ($\ell$
dependent) window of locality. Finally, we will argue that these zero modes
are a part of the full solution that decays for $r \gg L$, and therefore
solve the original equation as well.

%
%

\subsection{Equations for the zero modes}

We start from Eqs.~(\ref{eq:dtC1}) and (\ref{eq:Tab1}). In appendix
\ref{sec:simple} we perform integration by parts and algebraic
manipulations to bring the nontrivial integral in \Eq{eq:Tab1} to a more
tractable form. The result of this process is
\begin{eqnarray}
\label{eq:simpleT}
  && T^{\alpha\beta}(\B.r) = -\frac{1}{2}\PLO 
     K^{\mu\nu}\partial_\mu\partial_\nu C^{\alpha\beta}(\B.r) \nonumber \\
  && - \frac{1}{2} \frac{12\xi D}{(\xi-3)(\xi-5)}
      \int\!\! d\B.y \, G(\B.y) y^{\xi-2} 
         \partial^2 C^{\alpha\beta}(\B.r-\B.y) \ ,
\end{eqnarray}
which is true for every $\xi\neq 1$. The $\xi=1$ case will not be treated
here explicitly. Nevertheless, we will argue {\em a posteriori} that the
results for $\xi=1$ can be deduced from the $\xi\neq 1$ results by
continuity. 

Looking at \Eq{eq:simpleT}, we note that when $\xi=2$, the integral on the
RHS of the above equation trivializes to a local term
$C^{\alpha\beta}(\B.r)$. In this limiting case the model can be fully
solved utilizing the same machinery used in the previous section. The
solution can then be used to check the zero modes computed below for
arbitrary values of $\xi$.  

To proceed, we substitute \Eq{eq:simpleT} into \Eq{eq:dtC1}, noting that
the projector $\PRO$ leaves the nontrivial integral in \Eq{eq:simpleT}
invariant since it is divergence-free in both indices. Setting $\partial_t
C^{\alpha\beta}(\B.r,t) = 0$ in the stationary case, we arrive at the
following equation:
\begin{eqnarray}
\label{eq:fulleq}
 && 0 = -\Big[\PRO\PLO K^{\mu\nu}\partial_\mu\partial_\nu 
       C^{\alpha\beta}\Big](\B.r) \nonumber \\
 && -\frac{12\xi D}{(\xi-3)(\xi-5)}
      \int\!\! d\B.y\, G(\B.y) y^{\xi-2} \partial^2 
         C^{\alpha\beta}(\B.r-\B.y) \nonumber \\
 && + 2\kappa\partial^2 C^{\alpha\beta}(\B.r)+F^{\alpha\beta}(\B.r) \ .
\end{eqnarray}
As in the toy model, we apply two Laplacians to the above equation in order
to get rid of the integrals of the projection operators and obtain
\begin{eqnarray}
\label{eq:full-diff-eq}
 && 0 = -\partial^4\Big[\PRO\PLO K^{\mu\nu}\partial_\mu\partial_\nu 
       C^{\alpha\beta}\Big](\B.r) \nonumber \\
 && -\frac{12\xi D}{(\xi-3)(\xi-5)}
      \int\!\! d\B.y\, G(\B.y) y^{\xi-2} \partial^6
         C^{\alpha\beta}(\B.r-\B.y) \nonumber \\
 && + 2\kappa\partial^6 C^{\alpha\beta}(\B.r)+\partial^4 F^{\alpha\beta}(\B.r) \ .
\end{eqnarray}
Here and in the following, the operator $\partial^{2n}$ should be interpreted
as $(\partial^2)^n$. We now seek the homogeneous stationary solutions of
$C^{\alpha\beta}(\B.r)$ in the inertial range (zero modes). These satisfy
the equations obtained by neglecting the dissipation and setting the
forcing and time derivative to zero,
\begin{eqnarray}
\label{eq:zeroCab}
&& 0 = \partial^4
       K^{\mu\nu}\partial_\mu\partial_\nu C^{\alpha\beta}(\B.r)
   + \partial^\alpha\partial^\beta\partial_\tau\partial_\sigma 
       K^{\mu\nu}\partial_\mu\partial_\nu C^{\tau\sigma}(\B.r) \\
&&\quad -\ \partial^\alpha\partial_\tau\partial^2 
       K^{\mu\nu}\partial_\mu\partial_\nu C^{\tau\beta}(\B.r)
       - \partial^\beta\partial_\tau\partial^2 
       K^{\mu\nu}\partial_\mu\partial_\nu C^{\alpha\tau}(\B.r) \nonumber \\
&&\quad +\ \frac{12\xi D}{(\xi-3)(\xi-5)} 
   \int\!\! d\B.y \, G(\B.y) y^{\xi-2} 
     \partial^6 C^{\alpha\beta}(\B.r-\B.y) \ , \nonumber
\end{eqnarray}
Let us now define the RHS of the above equation as the ``zero-modes
operator'' $\OO(\xi)$ and write the zero-modes equation compactly as
\begin{equation}
  0 = \Big[ \OO(\xi) C^{\alpha\beta} \Big] (\B.r) \ .
\end{equation}
To solve it, we write the solution $C^{\alpha\beta}(\B.r)$ in a basis that
diagonalizes $\OO(\xi)$. This is done in the next subsection.

%
%

\subsection{The SO($3$) decomposition}

To diagonalize $\OO(\xi)$ we must look for its symmetries by looking for
the operations that commute with it. From \Eq{eq:zeroCab} it is easy to see
that these are rotations, scaling, permutation of indices, and flipping of
$\B.r$. As a result $\OO(\xi)$ is block diagonalized by tensors
$C^{\alpha\beta}(\B.r)$ that have the following properties:
\begin{itemize}
\item They belong to a definite sector $(\ell, m)$ of the SO($3$) group.
\item They have a definite scaling behavior, i.e., are proportional to
  $r^\lambda$ with some scaling exponent $\lambda$.
\item They are either symmetric or antisymmetric under permutations of
  indices.
\item They are either even or odd in $\B.r$.
\end{itemize}
In \cite{00ABP} we discuss these types of tensors in detail. Here we only
quote the final results. In every sector $(\ell, m)$ of the rotation group
with $\ell > 1$, one can find nine independent tensors $X^{\alpha\beta}(\B.r)$
that scale like $r^\lambda$. They are given by $r^\lambda B_{\ell m,
  q}^{\alpha\beta}(\hat{\B.r})$ where the index $q$ runs from 1 to 9
enumerating the different spherical tensors. These nine tensors can be
further subdivided into four subsets:
\begin{itemize}
\item \textbf{Subset I} of symmetric tensors with $(-)^\ell$ parity,
  containing 4 tensors.
\item \textbf{Subset II} of symmetric tensors with $(-)^{\ell+1}$ parity,
  containing 2 tensors.
\item \textbf{Subset III} of antisymmetric tensors with $(-)^{\ell+1}$ parity,
  containing 2 tensors.
\item \textbf{Subset IV} of symmetric tensors with $(-)^\ell$ parity,
  containing 1 tensor.
\end{itemize}
Due to the diagonalization of $\OO(\xi)$ by these subsets, the equation for
the zero modes foliates and we can compute the zero modes in each subset
separately. In this paper, we choose to focus on subset I, which has the
richest structure. The four tensors in this subset are given by
\begin{eqnarray}
  B_{1,\ell m}^{\alpha\beta}(\hat{\B.r}) &=& 
        r^{-\ell-2}r^\alpha r^\beta \Phi_{\ell m}(\B.r) \ , \\
  B_{2,\ell m}^{\alpha\beta}(\hat{\B.r}) &=& 
        r^{-\ell}[r^\alpha \partial^\beta + r^\beta\partial^\alpha] 
          \Phi_{\ell m}(\B.r) \ , \nonumber \\
  B_{3,\ell m}^{\alpha\beta}(\hat{\B.r}) &=& 
        r^{-\ell}\delta^{\alpha\beta} \Phi_{\ell m}(\B.r) \ , \nonumber \\
  B_{4,\ell m}^{\alpha\beta}(\hat{\B.r}) &=& 
        r^{-\ell+2}\partial^\alpha \partial^\beta \Phi_{\ell m}(\B.r) \ .
       \nonumber
\end{eqnarray}
We expect the calculation of the other subsets to be easier once one is
familiar with the techniques we are about to develop. Finally. we note that
since the correlation $C^{\alpha\beta}(\B.r)$ has to fulfill
$C^{\alpha\beta}(\B.r) = C^{\beta\alpha}(-\B.r)$, our subset I solution
will be valid only for even $\ell$s. 

Expanding $C^{\alpha\beta}(\B.r)$ in subset I,
\begin{eqnarray}
 C^{\alpha\beta}(\B.x) &=&  r^\lambda 
   \Big[ c_1 B_{1,\ell m}^{\alpha\beta}(\hat{\B.r}) + 
         c_2 B_{2,\ell m}^{\alpha\beta}(\hat{\B.r})  \nonumber \\
    &&+\ c_3 B_{3,\ell m}^{\alpha\beta}(\hat{\B.r}) +
         c_4 B_{4,\ell m}^{\alpha\beta}(\hat{\B.r}) \Big] \ , 
\label{eq:expand}
\end{eqnarray}
and plugging it back into PDE (\ref{eq:zeroCab}), we obtain a linear
equation for the coefficients $c_1, c_2, c_3, c_4$ that depend on the
parameters $\lambda, \ell, \xi$. In the four-dimensional space of column
vectors $(c_1, c_2, c_3, c_4)$, we can write it as
\begin{equation}
  \left( \begin{array}{ccc} 
      &&\\ \ \ \ & \OM(\lambda; \ell,\xi) & \ \ \ \\ && 
  \end{array}\right)
  \left( \begin{array}{c} c_1 \\ c_2 \\ c_3 \\ c_4 \end{array}\right) = 0 \ ,
\end{equation}
where $\OM(\lambda; \ell,\xi)$ is a $4 \times 4$ matrix that represents the
zero-modes operator $\OO(\xi)$. 

To continue, we note that due to the incompressibility constraint
(\ref{eq:C-Incomp}) of $C^{\alpha\beta}(\B.r)$, not all vectors $(c_1, c_2,
c_3, c_4)$ are allowed; for a given $\lambda, \ell$, only certain
combinations of the $B^{\alpha\beta}_{\ell m,q}(\hat{\B.r})$ lead to a
divergence free $C^{\alpha\beta}(\B.r)$. A simple calculation \cite{99ALP}
shows that these belong to a two-dimensional subspace, which is spanned by
the ``incompressible vectors''
\begin{eqnarray}
\label{eq:incomp-v12}
 && \Big|u_1(\lambda;\ell)\Big> = \left(\begin{array}{c}
     -\ell(\lambda-\ell) \\ (\lambda+2)(\ell-1)(\lambda-\ell+2)  \\
     0 \\ -(\lambda+3) \end{array} \right) \ , \nonumber \\
 && \Big|u_2(\lambda;\ell)\Big> = \left(\begin{array}{c}
     \lambda-\ell \\ 0 \\ (\lambda+2)(\ell-1)(\lambda-\ell+2) \\ 
     1 \end{array} \right) \ .
\end{eqnarray}
The zero modes exponents $\lambda$ can be found by requiring the equation
\begin{equation}
\label{eq:mat-zero}
  \OM(\lambda;\ell,\xi) \left[ a_1 \Big|u_1(\lambda; \ell)\Big> 
    + a_2 \Big|u_2(\lambda; \ell)\Big> \right] = 0\ ,
\end{equation}
to admit a nontrivial solution. The explicit computation of the matrix
$\OM(\lambda; \ell, \xi)$ is technically very cumbersome due to the
presence of the integral term. We shall therefore use an implicit method to
determine whether \Eq{eq:mat-zero} has a nontrivial solution or not. The
basic idea is that the calculation of the nontrivial integral in
\Eq{eq:zeroCab} can be simplified if we contract its free indices with the
two isotropic tensors $\hat{r}_\alpha \hat{r}_\beta$ and
$\delta_{\alpha\beta}$. Therefore, instead of solving \Eq{eq:zeroCab}
explicitly, we will contract its RHS with these two tensors and require
that the two resultant scalars vanish simultaneously. Obviously this would
provide us with a necessary condition for the solvability of
\Eq{eq:zeroCab}. Nevertheless, we shall see that it is also a sufficient
condition. 

Let us write the two tensors $\hat{r}_\alpha \hat{r}_\beta$ and
$\delta_{\alpha\beta}$ in matrix notation as two row vectors
$\Big<w_1(\ell)\Big|, \ \Big<w_2(\ell)\Big|$ given by
\begin{equation}
\begin{array}{lcll}
  \Big<w_1(\ell)\Big| &\equiv& 
     (\begin{array}{cccc} 1 & 2\ell & 1 & \ell(\ell-1) \end{array}) & 
     \mbox{contraction with $\hat{r}_\alpha \hat{r}_\beta$} \ , \\ \ \\
  \Big<w_2(\ell)\Big| &\equiv& 
       (\begin{array}{cccc} 1 & 2\ell & 3 & 0 \end{array}) & 
      \mbox{contraction with $\delta_{\alpha\beta}$} \ . 
\end{array}
\end{equation}
The contraction of $\delta_{\alpha\beta}$ and $\hat{r}_\alpha
\hat{r}_\beta$ with another tensor is translated to the usual matrix
multiplication of these row vectors with a column vector. For example, if
$\Big | c \Big>$ is a column vector whose components are given by $c_i,\ 
i=1,\ldots 4$, then
\begin{eqnarray}
&& \delta_{\alpha\beta}\Big( c_1 B^{\alpha\beta}_{1,\ell m}(\hat{\B.r})
   + c_2 B^{\alpha\beta}_{2,\ell m}(\hat{\B.r}) 
   + c_3 B^{\alpha\beta}_{3,\ell m}(\hat{\B.r})
   + c_4 B^{\alpha\beta}_{4,\ell m}(\hat{\B.r}) \Big) \nonumber \\ 
&&\quad =  \Big(c_1 + 2\ell c_2 + 3 c_3\Big)Y_{\ell m}(\hat{\B.r})
             \nonumber \\
&&\quad =  \Big< w_2(\ell) \Big| c \Big>Y_{\ell m}(\hat{\B.r}) \ .
\label{eq:kat-bra}
\end{eqnarray}

Returning to the zero-mode equations, we now define the $2\times 2$
``reduced matrix'' $O_{ij}(\lambda; \ell, \xi)$ by
\begin{equation}
  O_{ij}(\lambda; \ell, \xi) \equiv 
    \Big< w_i(\ell) \Big| \OM(\lambda; \ell, \xi) 
       \Big| u_j(\lambda; \ell) \Big> \ , \label{redmat}
\end{equation}
Obviously, the existence of a nontrivial solution that makes the two
contractions vanish is equivalent to the requirement that $O_{ij}(\lambda;
\ell, \xi)$ is singular, i.e.,
\begin{equation}
\label{eq:solvability}
  \det O_{ij}(\lambda; \ell, \xi) = 0 \ .
\end{equation}
The above condition is also sufficient for the solvability of the
zero-modes equation. To see that, notice that the RHS of \Eq{eq:fulleq} and
therefore the RHS of \Eq{eq:zeroCab} produce tensors that are
divergence-free in both indices. Thus the vectors $\OM(\lambda; \ell,
\xi)\Big| u_i(\lambda; \ell)\Big>$ will belong to the two-dimensional
subspace that is spanned by $\Big|u_j(\lambda+\xi-6;\ell)\Big>$. Since the
transformation matrix
\begin{equation}
  U_{ij} \equiv  \Big< w_i(\ell) \Big| u_j(\lambda+\xi-6; \ell) \Big> \ ,
\end{equation}
is nonsingular for all values of $\lambda$ and $\xi$, we find that
$\OM(\lambda; \ell, \xi)\Big| u_1(\lambda; \ell)\Big>$ and $\OM(\lambda;
\ell, \xi)\Big| u_2(\lambda; \ell)\Big>$ are linearly dependent if and only
if \Eq{eq:solvability} holds.

Looking at \Eq{eq:zeroCab}, we recognize that $\OO(\xi)$ is a sum of a
differential and an integral operator. Consequently, $\OM(\lambda;
\ell,\xi)$ and $O_{ij}(\lambda; \ell, \xi)$ can be written as a sum of
corresponding parts,
\begin{eqnarray}
  \OM(\lambda; \ell, \xi) &=& \OM^{dif}(\lambda; \ell, \xi) + 
     \OM^{int}(\lambda; \ell, \xi) \ , \\
  O_{ij}(\lambda; \ell, \xi) &=& O^{dif}_{ij}(\lambda; \ell, \xi) + 
     O^{int}_{ij}(\lambda; \ell, \xi) \ . \nonumber
\end{eqnarray}
These parts are of a different nature and thus we dedicate two different
subsections for their calculation.
\end{multicols}
\sepBR{8.6}{-0.3}{.5}

%
%

\subsubsection{Form of the differential operator}
The calculation of $O_{ij}^{dif}(\lambda; \ell,\xi)$ can be done directly
by calculating the matrix $\OM^{dif}(\lambda; \ell, \xi)$, employing the
same techniques that are used in Refs \cite{99ALP,00ABP} and in
sec.~\ref{sec:Toy}. Here we merely give the mid and final results. 

From \Eq{eq:zeroCab}, we see that the differential part of the zero-mode
operator is given by
\begin{eqnarray}
\OO^{dif}(\xi) C^{\alpha\beta} &\equiv&
  \partial^2\partial^2 
       K^{\mu\nu}\partial_\mu\partial_\nu C^{\alpha\beta}
       + \partial^\alpha\partial^\beta\partial_\tau\partial_\sigma
       K^{\mu\nu}\partial_\mu\partial_\nu C^{\tau\sigma} \\
 &&- \partial^\alpha\partial_\tau\partial^2 
       K^{\mu\nu}\partial_\mu\partial_\nu C^{\tau\beta}
       - \partial^\beta\partial_\tau\partial^2 
       K^{\mu\nu}\partial_\mu\partial_\nu C^{\alpha\tau} \nonumber \ .
\end{eqnarray}
To find its matrix representation it is convenient to first calculate
following operators:
\begin{itemize}
\item $\LM(\lambda;\ell): 
  X^{\alpha\beta} \longrightarrow \partial^2 X^{\alpha\beta}$
\begin{equation}
  \LM(\lambda;\ell) = \left(\begin{array}{llll}
     (\lambda-\ell-2)(\lambda+j+3) & 0 & 0 & 0 \\ 
     2 & (\lambda-\ell)(\lambda+\ell+1) & 0 & 0 \\ 
     2 & 0 & (\lambda-\ell)(\lambda+\ell+1) & 0 \\ 
     0 & 4 & 0 & (\lambda-\ell+2)(\lambda+\ell-1) \end{array} \right) \ ,
\end{equation}
\item $\MM_1(\lambda;\ell): X^{\alpha\beta} \longrightarrow 
   \partial^\alpha\partial^\beta\partial_\tau\partial_\sigma X^{\tau\sigma}$
\begin{equation}
\MM_1(\lambda;\ell) = 
 \left(\begin{array}{c} (\lambda-\ell-2)(\lambda-\ell-4) \\ 
      \lambda - \ell -2 \\ \lambda-\ell-2 \\ 1 \end{array} \right) \times
 \left(\begin{array}{c}  (\lambda+1) (\lambda+2)  \\
      2\ell(\lambda+2)(\lambda-\ell) \\ (\lambda+1+\ell)(\lambda-\ell) \\
      \ell(\lambda-\ell)(\ell-1)(\lambda-\ell+2) \end{array} 
   \right)^{\mbox{\large T}}
\end{equation}
\item $\MM_2(\lambda;\ell): X^{\alpha\beta} \longrightarrow 
   \partial^\alpha\partial_\tau X^{\tau\beta} + 
   \partial^\beta\partial_\tau X^{\alpha\tau} $
\begin{equation}
\MM_2(\lambda;\ell) = 
  \left(\begin{array}{llll}
    2(\lambda-\ell-2)(\lambda+2)  & 2\ell(\lambda-\ell)(\lambda-\ell-2) 
    & 2(\lambda-\ell-2)(\lambda-\ell) & 0 \\ 
    \lambda+2 & (\lambda-\ell)(\ell+\lambda+3) & 2(\lambda-\ell) 
    & (\lambda-\ell)(\ell-1)(\lambda-\ell+2) \\ 
    2(\lambda+2) & 2\ell(\lambda-\ell) & 2(\lambda-\ell) & 0 \\ 
    0 & 2(\lambda+3) & 2 & 2(\ell-1)(\lambda-\ell+2) 
\end{array}\right)  \ ,
\end{equation}
\item $\KM(\lambda;\ell,\xi): X^{\alpha\beta} \longrightarrow 
   K^{\mu\nu}(\xi)\partial_\mu\partial_\nu X^{\alpha\beta}$
\begin{equation}
\KM(\lambda;\ell, \xi) = 
  D\Big[ (\xi+2)\LM(\lambda; \ell) - \xi\lambda(\lambda-1)\openone \Big]
\end{equation}
\end{itemize}
\begin{multicols}{2}
With these four matrices, the matrix form of the differential part is
written compactly as
\begin{eqnarray}
&&  \OM^{dif}(\lambda; \ell, \xi) =
   \Big[\LM(\lambda+\xi-4; \ell)\LM(\lambda+\xi-2; \ell) \\ 
  &&\quad + \ \MM_1(\lambda+\xi-2; \ell) \nonumber \\
  &&\quad - \ \MM_2(\lambda+\xi-4; \ell)\LM(\lambda+\xi-2; \ell) \Big]
    \KM(\lambda; \ell, \xi) \ .
\nonumber
\end{eqnarray}
$O_{ij}^{dif}(\lambda; \ell,\xi)$ is then computed directly from
definition.
%
%
\subsubsection{Form of the integral operator}
The integral part of the zero-modes operator is given by
\begin{eqnarray}
  &&[\OO^{int}(\xi) C^{\alpha\beta}](\B.r) \equiv \nonumber\\
   && \frac{12\xi D}{(\xi-3)(\xi-5)} 
     \int\!\! d\B.y \, G(\B.r-\B.y) |\B.r-\B.y|^{\xi-2} 
       \partial^6 C^{\alpha\beta}(\B.y) \ . 
\end{eqnarray}
To calculate the reduced matrix (\ref{redmat}), we have to compute the
contraction of $\OO^{int}(\xi)C^{\alpha\beta}$ with $\hat{r}_\alpha
\hat{r}_\beta$ and $\delta_{\alpha\beta}$. To do that, let the tensor
$X^{\alpha\beta}(\B.r)$ be of the form of the trial solution
(\ref{eq:expand}), i.e., let it belong to the $(\ell, m)$ sector of
SO($3$), be divergence free in both indices, and proportional to
$r^\lambda$. If we denote in matrix notation $X^{\alpha\beta}(\B.r)$ by the
column vector $\Big| x\Big>$, then from the isotropy of the tensors
$\hat{r}_\alpha \hat{r}_\beta, \ \delta_{\alpha\beta}$, we have the
following useful identities:
\begin{eqnarray}
\label{eq:cont}
  \hat{r}_\alpha \hat{r}_\beta X^{\alpha\beta}(\B.r) &=& 
    \Big< w_1(\ell) \Big| x \Big> r^{\lambda}Y_{\ell m}(\hat{\B.r}) \ , \\
  \delta_{\alpha\beta} X^{\alpha\beta}(\B.r) &=& 
    \Big< w_2(\ell) \Big| x \Big> r^{\lambda}Y_{\ell m}(\hat{\B.r}) \ .
\nonumber
\end{eqnarray}

Consider now the contraction of
\begin{equation}
   \int\!\! d\B.y \, G(\B.r-\B.y) |\B.r-\B.y|^{\xi-2} 
      X^{\alpha\beta}(\B.y) \ ,
\end{equation}
with $\delta_{\alpha\beta}$
\begin{eqnarray}
&&I_1 =  \delta_{\alpha\beta} \int\!\! d\B.y \, G(\B.r-\B.y) 
       |\B.r-\B.y|^{\xi-2} X^{\alpha\beta}(\B.y) \\ 
   && \quad = \Big< w_2(\ell) \Big| x \Big> 
      \int\!\! d\B.y \, G(\B.r-\B.y) |\B.r-\B.y|^{\xi-2} 
        y^\lambda Y_{\ell m}(\hat{\B.y}) \ .
\nonumber
\end{eqnarray}
If $\lambda$ is in the window of locality of the integral above, then the
integral will converge and be proportional to $r^{\lambda+\xi}Y_{\ell
  m}(\hat{\B.r})$. The scaling exponent $\lambda+\xi$ follows from power
counting [remember that $G(\B.r) \sim 1/r$] while the angular dependence
is a result of the isotropy of the integration over all space.  This leads
us to define the proportionality factor $A(\lambda; \xi, \ell)$,
\begin{eqnarray}
\label{def:A}
&& A(\lambda; \ell, \xi) r^{\lambda+\xi} Y_{\ell m}(\hat{\B.r}) \\
&& \quad \equiv \ \int\!\! d\B.y \, G(\B.r-\B.y) 
          |\B.r-\B.y|^{\xi-2} y^\lambda Y_{\ell m}(\hat{\B.y}) \nonumber \ ,
\end{eqnarray}
with which we can write
\begin{equation}
\label{eq:I1}
  I_1 = \Big< w_2(\ell) \Big| x \Big> 
    A(\lambda; \ell, \xi) r^{\lambda+\xi} Y_{\ell m}(\hat{\B.r}) \ .
\end{equation}

The prefactor $A(\lambda; \ell, \xi)$ is the only part of the calculation
that cannot be done analytically. However, instead of calculating the
integral in \Eq{def:A} numerically, we can expand the integrand as a
Taylor series and write $A(\lambda; \ell, \xi)$ as an infinite series of
poles in $\lambda$. This is done in detail in Appendix \ref{sec:A} with the
result that
\begin{eqnarray}
\label{eq:A}
&& A(\lambda; \ell, \xi) = \\
&& -\frac{1}{2} \sum_{q=0}^{\infty} a_q(\ell,\xi) 
   \left[ \frac{1}{\lambda+3+\ell+2q} - 
     \frac{1}{\lambda+\xi-\ell-2q}\right] \ .
\nonumber
\end{eqnarray}
The coefficients $ a_q(\ell,\xi)$ are given in Appendix \ref{sec:A}. We
notice that the window of locality in the definition (\ref{def:A}) of
$A(\lambda; \xi, \ell)$ can be identified from the positions of the poles
in \Eq{eq:A}. Indeed, the boundaries of the window of locality are
determined by the $q=0$ term and located at $\lambda=-\ell-3$ and
$\lambda=\ell-\xi$. When $\lambda$ hits these boundaries the integral in
\Eq{def:A} diverges corresponding to a pole in \ref{eq:A}. However, the
above formula is valid also for values of $\lambda$ outside this window of
locality for which the formal definition of $A(\lambda; \ell, \xi)$ makes
no sense. The relevance of these values to the full solution must therefore
be addressed. This will be done in the next subsection where we present the
values of the scaling exponents. 

Let us now consider the $\hat{r}_\alpha \hat{r}_\beta$ contraction
\begin{eqnarray}
&&I_2 =  r^{-2}\int\!\! d\B.y \, G(\B.r-\B.y) 
       |\B.r-\B.y|^{\xi-2} r_\alpha r_\beta X^{\alpha\beta}(\B.y) \ .  
\end{eqnarray}
Using the identity
\begin{eqnarray}
&& r_\alpha r_\beta = (r_\alpha-y_\alpha)(r_\beta-y_\beta) 
    + (r_\alpha - y_\alpha)y_\beta \\ 
&& \quad +\  y_\alpha(r_\beta-y_\beta)  + y_\alpha y_\beta \ ,
\nonumber
\end{eqnarray}
we can decompose the integral into four terms:
\begin{itemize}
\item The $(r_\alpha-y_\alpha)(r_\beta-y_\beta)$ term: \\ \ \\
  Recalling that
  \begin{equation}
    G(\B.r-\B.y) = -\frac{1}{4\pi|\B.r-\B.y|} \ ,
  \end{equation}
  it is easy to verify that
  \begin{eqnarray}
    &&  G(\B.r-\B.y)|\B.r-\B.y|^{\xi-2}(r_\alpha-y_\alpha)(r_\beta-y_\beta) \\
    &&\quad = \frac{1}{(\xi+1)(\xi-1)}\partial_\alpha\partial_\beta 
          G(\B.r-\B.y)|\B.r-\B.y|^{\xi+2} \nonumber \\
    && \quad - \ \frac{1}{\xi-1}G(\B.r-\B.y)|\B.r-\B.y|^\xi 
         \delta_{\alpha\beta} \ . \nonumber
  \end{eqnarray}
  Plugging this into the integral, the term with the derivatives will
  vanish due to integration by parts and the divergence-free
  $X^{\alpha\beta}(\B.r)$. We are therefore left with
  \begin{eqnarray}
    && -\frac{r^{-2}}{\xi-1} \int\!\! d\B.y \, G(\B.r-\B.y)|\B.r-\B.y|^\xi 
      \delta_{\alpha\beta} X^{\alpha\beta}_{\ell m}(\B.y) \\
    && = -\frac{r^{-2}}{\xi-1}\Big<w_2(\ell)\Big| x\Big>  
\!\!\!\int\!\! d\B.y \, G(\B.r-\B.y)|\B.r-\B.y|^\xi 
      y^\lambda Y_{\ell m}(\hat{\B.y}) \ . \nonumber
  \end{eqnarray}
  Using the identity
  \begin{equation}
    \partial^2 y^{\lambda+2}Y_{\ell m}(\hat{\B.y}) =
      (\lambda+2-\ell)(\lambda+3+\ell) y^\lambda Y_{\ell m}(\hat{\B.y}) \ ,
  \end{equation}
  we further integrate by parts the last integral to finally obtain
  \begin{eqnarray}
    \nonumber
    && \frac{-\xi}{(\lambda+2-\ell)(\lambda+3+\ell)}
         \Big<w_2(\ell)\Big| x\Big> A(\lambda+2; \ell, \xi)  \\
    &&\quad \times \ r^{\lambda+\xi}Y_{\ell m}(\hat{\B.r})  \ . 
  \end{eqnarray}
\item The $(r_\alpha - y_\alpha)y_\beta, \ y_\alpha(r_\beta - y_\beta)$ terms:
      \\ \ \\
      Using the same tricks as in the previous term, i.e., integration by
      parts and the fact that $X^{\alpha\beta}(\B.y)$ is divergence free,
      we can easily show that both these terms vanish.
\item The $y_\alpha y_\beta$ term:
      \\ \ \\
      Using \Eq{eq:cont} and the definition of $A(\lambda; \ell, \xi)$, we
      directly get
  \begin{eqnarray}
     \Big<w_1(\ell)\Big| x\Big> A(\lambda+2; \ell, \xi)
       r^{\lambda+\xi}Y_{\ell m}(\hat{\B.r})  \ . 
  \end{eqnarray}
\end{itemize}
Gathering all the terms, we see that the contraction with $\hat{r}_\alpha
\hat{r}_\beta$ is equal to
\begin{eqnarray}
&& I_2 = \left[ \Big<w_1(\ell)\Big| x\Big> - 
    \frac{\xi}{(\lambda+2-\ell)(\lambda+3+\ell)}
    \Big<w_2(\ell)\Big| x\Big> \right] \nonumber \\
&& \quad\quad \times \ A(\lambda+2; \ell, \xi)
    r^{\lambda+\xi}Y_{\ell m}(\hat{\B.r})  \ . 
\label{eq:I2}
\end{eqnarray}
To conclude, the matrix $O^{int}_{ij}(\lambda; \ell, \xi)$ is given by the
following equations:
\end{multicols}
\sepBR{8.6}{0}{1}
\begin{eqnarray}
&& \Big| \tilde{u}_i(\lambda; \ell, \xi) \Big> = 
    \frac{12\xi D}{(\xi-3)(\xi-5)} 
    \LM(\lambda-4; \ell)\LM(\lambda-2; \ell)\LM(\lambda; \ell)
    \Big| u_i(\lambda; \ell) \Big>  \ , \\
&&  O^{int}_{1,i}(\lambda; \ell, \xi) = A(\lambda-4; \ell,\xi) \left[ 
   \Big<w_1(\ell)\Big| \tilde{u}_i(\lambda; \ell, \xi) \Big> - 
    \frac{\xi}{(\lambda-4-\ell)(\lambda-3+\ell)}
    \Big<w_2(\ell)\Big| \tilde{u}_i(\lambda; \ell, \xi)\Big> 
      \right] \ , \nonumber \\
&&  O^{int}_{2,i}(\lambda; \ell, \xi) =  A(\lambda-6; \ell, \xi) 
   \Big<w_2(\ell)\Big| \tilde{u}_i(\lambda; \ell, \xi) \Big> \ .
\nonumber
\end{eqnarray}
\rightline{\sepTL{8.7}{0.4}{0.5}}
\begin{multicols}{2}

%
%

\subsection {Results}  

Figure 2 shows the leading scaling exponents of the linear pressure model
for $\ell=0,2,4,6,8,10$. The results where obtained by numerically solving
\Eq{eq:solvability} for $\xi=0,0.01, 0.02, \ldots, 1.99,2$. The prefactor
$A(\lambda; \ell, \xi)$ was calculated using the formulas
(\ref{eq:A-final}) and (\ref{def:aq}) where the infinite series of poles
was truncated typically after 100 poles, when it was clear that relative
contribution from the consecutive pole was smaller than the machine
precision (about $10^{-14}$). Additionally, the numerical results were
compared to the analytical results of $\xi=0,2$ and of $\ell=0$ and
$\ell=2$ and found to be correct within a relative error of $10^{-10}$.

From Fig.~2, we see that in the isotropic sector and in the $\ell=2$
sector, the leading exponent is $\zeta = 0$, corresponding to the trivial
$C^{\alpha\beta}(\B.r)=\mbox{const}$ solution. These zero modes will not
contribute to the second-order structure function, which is given by
\begin{equation}
  S^{\alpha\beta}(\B.r) = 
     2\Big[C^{\alpha\beta}(\B.r) - C^{\alpha\beta}(\B.0)\Big]
\end{equation}
and so we have to consider the zero mode with the consecutive exponent. In
the isotropic sector this exponent is exactly $\zeta=2-\xi$ as can be
proven by passing to Fourier space. This special solution is a finger-print
of the existence of a constant energy flux in this model. Indeed just like
in Navier-Stokes turbulence, one can show analytically that the isotropic
part of the triple correlation function $\Big< v^\alpha(\B.x)
w^\mu(\B.x+\B.r) v^\beta(\B.x+\B.r)\Big>$ is proprtional to $r$ and hence
$S^{\alpha\beta}_{(\ell=0)}(\B.r) \sim r^{2-\xi}$.

Returning to the main question of this paper, we see that no saturation of
the anisotropic exponents occurs since the leading exponent in every
$\ell>2$ sector is $\zeta^{(\ell)} \simeq \ell-2$. These exponents are
within the window of locality of \Eq{eq:fulleq}, which is given by $-\ell-3
< \zeta < \ell - \xi$. However, the next-to-leading exponents (that are the
leading ones in the structure function for $\ell=0,2$) are already out of
this window and their relevance has to be discussed. We propose that the
same mechanism that works in the toy model (see sec.~\ref{sec:toy-bc})
also operates here and that all these higher exponents can be found in the
full solution. To understand this, let us write a model equation for the
correlation function in the spirit of Eq.(\ref{eq:ex}),
\begin{equation}
  \hat{\cal D}C(\B.r) + \int d\B.y\, K(\B.r-\B.y) C(\B.y) = F(\B.r)
\label{eq:model}
\end{equation}
with $K$ being some kernel, and $\hat{\cal D}$ being some local
differential operator. In view of \Eq{eq:fulleq}, the differential operator
$\hat{\cal D}$ should be regarded as the Kraichnan operator and the
integral term should be taken for all integral terms in the equation
including integrals due to the projection operators. These integrals create
a window of locality that we denote by $\lambda_{\mbox{\small low}} <
\lambda < \lambda_{\mbox{\small hi}}$. Any pure scaling solution $C(\B.r)
\sim r^\lambda$ with $\lambda$ outside the window of locality will diverge
and hence will not solve the homogeneous part of \Eq{eq:model}.
Nevertheless, we will now demonstrate how this zero mode can be a part of a
full solution without breaking scale invariance. For this we act with a
Laplacian on both sides of \Eq{eq:model} in order to get rid of the
projection operators integrals.  Of course, like in the linear pressure
model, this will not eliminate all integral terms, and thus we can write
the resultant equation as
\begin{equation}
  \partial^2\hat{\cal D}C(\B.r) 
     + \int d\B.y \, K(\B.r-\B.y)\partial^2 C(\B.y) = \partial^2 F(\B.r) \ .
\label{eq:lap-model}
\end{equation}
Our main assumption, which was proven analytically in the simple case of
the toy model, is that the above equation has a solution that is finite for
all $r$ and decays for $r \gg L$. Let us now consider the zero modes of
\Eq{eq:lap-model}; their exponents have to be within the ``shifted'' window
of locality $\lambda_{\mbox{\small low}}+2 < \lambda <
\lambda_{\mbox{\small hi}}+2$. Suppose now that $r^\lambda$ with
$\lambda_{\mbox{\small hi}} < \lambda < \lambda_{\mbox{\small hi}}+2$ is
such a solution, which is therefore part of the full solution of
\Eq{eq:lap-model}. We now claim that this solution also solves the original
equation [\Eq{eq:model}], hence allowing the existence of scaling exponents
outside its window of locality. To see that, we first notice that since the
full solution decays for $r\gg L$, then all the integrals in \Eq{eq:model}
converge and are therefore well defined. All that is left to show is that
the equation is indeed solved by $C(\B.r)$. But this is a trivial
consequence of the uniqueness of the solution for Laplace equation with
zero at infinity boundary conditions. Indeed, if we denote the integral
term in \Eq{eq:model} by
\begin{equation}
  I(\B.r) = \int d\B.y\, K(\B.r-\B.y) C(\B.y) \ ,
\end{equation}
then from \Eq{eq:lap-model} we have
\begin{equation}
  \partial^2 I(\B.r) = \partial^2 [F(\B.r) - \hat{\cal D}C(\B.r)] \ ,
\end{equation}
and since both $I(\B.r)$ and $F(\B.r) - \hat{\cal D}C(\B.r)$ decay as
$r\to\infty$, then they must be equal. Of course no breaking of scale
invariance occurs because the equation is satisfied and $F(\B.r) - \hat{\cal
  D}C(\B.r)$ is a sum of an inhomogeneous solution and power laws.

Returning to the linear pressure model, we have shown that not only the
first, leading exponents in every sector are legitimate but also the next
few exponents. These exponents are inside the shifted window of locality of
the ``Laplaced'' equation (\ref{eq:zeroCab}), which is given by $-\ell+1 <
\lambda < \ell+4-\xi$.

At this point, we may ask whether this is also the case for the other
exponents, which are outside this shifted window of locality. In light of
the above discussion, it is clear that all of them may also be part of the
full solution for we can always differentiate \Eq{eq:fulleq} sufficient
number of times, thus shifting the window of locality to include any of
these exponents. However, this procedure is unnecessary once we have
written the prefactor $A(\lambda; \ell, \xi)$ as an infinite sum of poles
in $\lambda$.  In that case the equation is defined for all values of
$\lambda$ except for a discrete set of poles, enabling us to look for
exponents as high as we wish.
\end{multicols}
\begin{figure}[h]
\label{fig:linearP-exponents}
\epsfxsize=18cm
\epsfbox{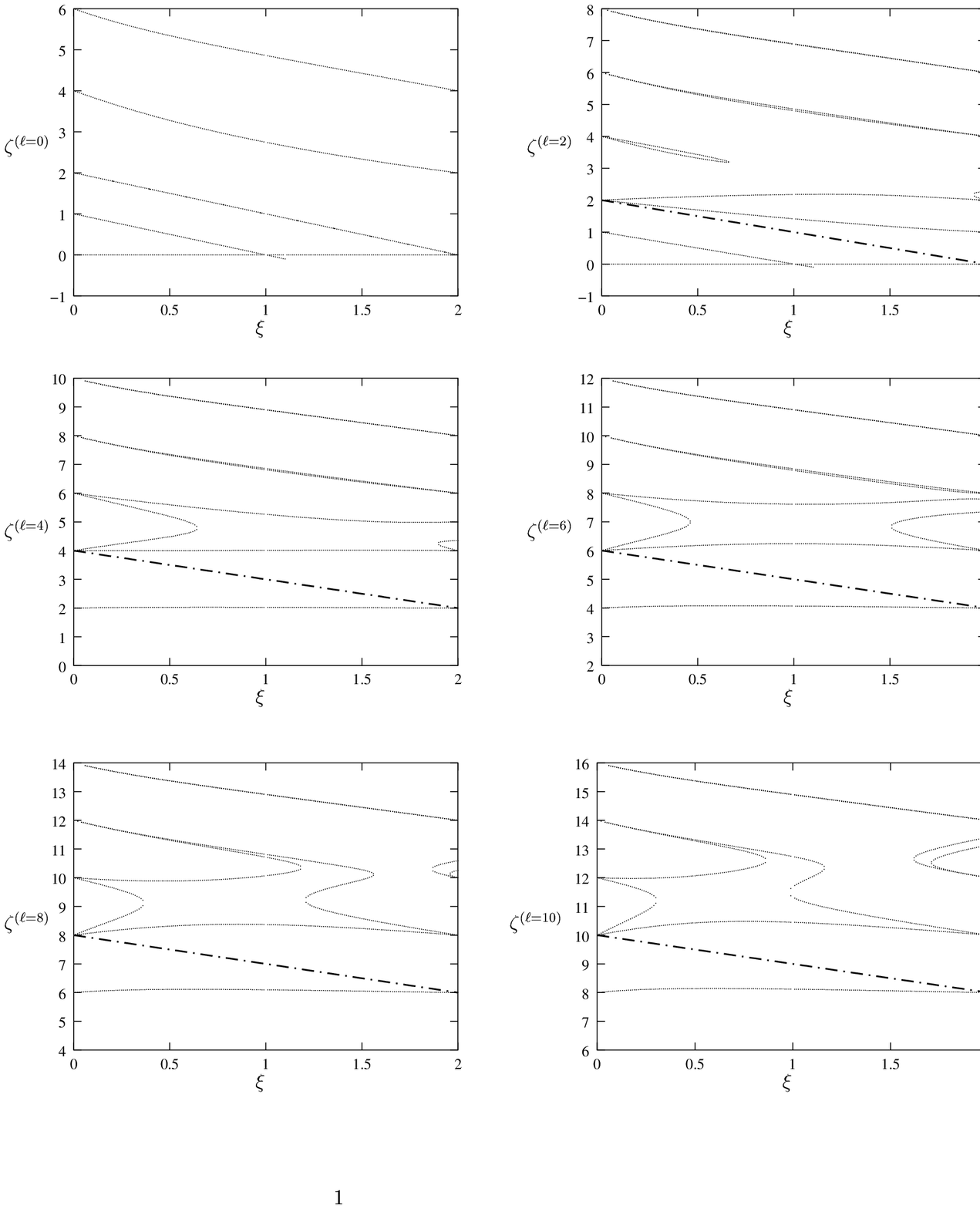}
\caption{Leading scaling exponents for the first few $\ell$'s in the Linear
  Pressure model. The dashed line indicates the upper bound of the window
  of locality}
\end{figure}
\begin{multicols}{2}

%
%

\section{Summary and conclusions}
\label{sec:summary}

The main question raised and answered in this paper is whether the
existence of the pressure terms necessarily leads to a saturation of the
scaling exponents associated with the anisotropic sectors. Such terms
involve integrals over all space, and seem to rule out the existence of an
unbounded spectrum. We have discussed a mechanism that allows an unbounded
spectrum without spoiling the convergence of the pressure integrals. The
mechanism is demonstrated fully in the context of the simple toy model and
we proposed that it also operates in the case of the linear pressure model.
The mechanism is based on two fundamental observations.  The first one is
that the window of locality widens up linearly in $\ell$ due to the angular
integration. The second, and more important, is that a scaling solution
with an unbounded spectrum can exist {\em as a part of a full solution,
  which decays at infinity}. Indeed pure scaling solutions cannot solve by
themselves the zero-modes equation if their scaling exponent is out of the
window of locality. However, the zero-modes are always part of the full
solution that decays to zero once $r \gg L$ and we have shown that if
such a solution solves a differential version of the full equation, it must
also solve the original equation. Therefore by differentiating the full
equation sufficiently many times, we can always reach a differential
equation with a window of locality as high as we wish. In that equation we
can find zero-mode solutions with arbitrarily high exponents (notice that
in the toy model, it was sufficient to differentiate once to get rid of all
integrals, thus obtaining an ``infinitely wide'' window of locality). But
since these zero-modes are part of a full solution that decays at infinity,
then this solution is also valid for the original equation, hence showing
that in the full solution there can be power laws with arbitrarily high
exponents.

Finally, we want to comment about the relevance of our calculations to
Navier-Stokes turbulence. If we substitute blindly $\xi=4/3$ in our
results, we predict the exponents 2/3, 1.252\ 26, 2.019\ 22, 4.048\ 43,
6.068\ 60 and 8.083\ 37 for $\ell=0,2,4,6,8$, and 10, respectively. It
would be tempting to propose that similar numbers may be expected for
Navier-Stokes and indeed for $\ell=0$ and 2 this is not too far from the
truth. We cannot, however, state with confidence that the genuine
nonlinearity of Navier-Stokes does not change these numbers significantly.
More work is needed before we can draw final conclusions on the rate of
decay of the high sectors of anisotropy in Navier-Stokes turbulence.

%
%

\acknowledgments 

We thank Luca Biferale, Yoram Cohen and Massimo Vergassola for many helpful
discussions and suggestions. We would also like to thank Anna Pomyalov for
her help in the numerical calculations. This work has been supported in
part by the Israel Science Foundation, the German-Israeli Foundation, the
European Commission under contract No. HPRN-CT-2000-00162 (``Nonideal
Turbulence"), and the Naftali and Anna Backenroth-Bronicki Fund for
Research in Chaos and Complexity.

\appendix

%
%

\section{Gaussian integration by parts}
\label{sec:gauss}

The field $\B.{w}(\B.x,t)$ as well as the forcing are Gaussian white
noises. This enables us to express $T^{\alpha\beta}(\B.r)$ and the
correlation of the force in terms of $C^{\alpha\beta}(\B.r)$ and
$F^{\alpha\beta}(\B.r)$. One way to accomplish this is by using the
Gaussian integration by parts method \cite{ref:GI}. Using the basic formula
for Gaussian integration by parts, we get for the third moment
\begin{eqnarray}
\label{eq:gauss}
&&  \left<v^\alpha(\B.x+\B.r,t) w^\mu(\B.x,t)v^\beta(\B.x,t) \right> = \\
&& \quad \int\!\! dt' \int \!\!\, d\B.y\,
  \left<w^\mu(\B.x,t) w^\nu(\B.y,t') \right> \nonumber \\
  && \quad \times \Big[
  \left<\frac{\delta v^\alpha(\B.x+\B.r,t)}{\delta w^\nu(\B.y,t')}
    v^\beta(\B.x,t)\right>  \nonumber \\
  && \quad + \ \left<v^\alpha(\B.x+\B.r,t)
    \frac{\delta v^\beta(\B.x,t)}{\delta w^\nu(\B.y,t')}\right> \Big] \ .
\nonumber
\end{eqnarray}
To find out the functional derivative, we formally integrate
$v^\alpha(\B.x,t)$
\begin{eqnarray}
&&v^\alpha(\B.x,t) 
  = \int_{-\infty}^t \!\! dt'\, \partial_{t'} v^\alpha(\B.x,t') \nonumber \\
&& =\ -\int_{-\infty}^t \!\!dt' \, 
     w(\B.x,t')^\mu\partial_\mu v^\alpha(\B.x,t') \nonumber \\
&&+\ \int_{-\infty}^t \!\!dt' \!\! \int\!\! d\B.y \, 
    [\partial^\alpha\partial_\tau G(\B.x-\B.y)] 
       w^\mu(\B.y,t') \partial_\mu v^\tau(\B.y,t') \nonumber \\ 
&& + \ [\mbox{terms that are independent of $\B.w$}] \ ,
\end{eqnarray}
and thus
\begin{eqnarray}
\label{eq:func-derivative}
&&  \frac{\delta v^\alpha(\B.x,t)}{\delta w^\nu(\B.y,t')}
  = \theta(t-t')\Big[ -\delta^3(\B.x-\B.y)\partial_\nu v^\alpha(\B.y,t') \\
&& \quad + \ [\partial^\alpha\partial_\tau G(\B.x-\B.y)] 
       \partial_\nu v^\tau(\B.y,t') \Big] \ .
\nonumber
\end{eqnarray}
When we plug this result back to \Eq{eq:gauss} we face the problem of
evaluating the step function $\theta(t-t')$ at $t=t'$ due to the delta
correlation in time of $\B.w(\B.r, t)$. To solve this problem in a
``physical'' way \cite{ref:GI}, we approximate the delta function of the
white noise with a sharp even function, perform the integral, and only then
take the white noise limit. Doing so we obtain the formal result
$\theta(0)=1/2$ stemming from the fact that we approximate a delta
function with an even function. Finally, we remark that this derivation
corresponds to the Stratonovich interpretation of the stochasic equation
\Eq{eq:v-p}.

Next, we perform the spatial integration, arriving at
\begin{eqnarray}
  &&\left<v^\alpha(\B.x+\B.r,t)w^\mu(\B.x,t)v^\beta(\B.x,t) \right>
  \nonumber \\ 
 &=& -\frac{1}{2} K^{\mu\nu}(\B.r)\partial_\nu 
       C^{\alpha\beta}(\B.r) 
     + \frac{1}{2}\partial_{(r)}^\alpha \nonumber \\
 && \times \int\!\! d\B.y\, G(\B.r-\B.y)\partial_\tau\Big[K^{\mu\nu}(\B.y)
         \partial_\nu C^{\tau\beta}(\B.y)\Big] \nonumber \\
 && +\  \frac{1}{2}\int\!\!d\B.y\, \partial^\beta\partial_\tau G(\B.y) 
      \Big[K^{\mu\nu}(\B.y)\partial^{(y)}_\nu 
          C^{\alpha\tau}(\B.r-\B.y)\Big] \ ,
\end{eqnarray}
and therefore
\begin{eqnarray}
&&  T^{\alpha\beta}(\B.r) = \partial^{(r)}_\mu 
    \left<v^\alpha(\B.x+\B.r)w^\mu(\B.x)v^\beta(\B.x)\right> \\
&& \quad =  -\frac{1}{2}K^{\mu\nu}(\B.r)\partial_\mu\partial_\nu 
   C^{\alpha\beta}(\B.r) \nonumber \\
&& \quad +\  \frac{1}{2}\partial^\alpha_{(r)}
  \int\!\!d\B.y \, G(\B.r-\B.y)\partial_\tau\Big[K^{\mu\nu}(\B.y)
         \partial_\mu\partial_\nu C^{\tau\beta}(\B.y)\Big] \nonumber \\
&& \quad -\ \frac{1}{2}\int\!\! d\B.y \, 
      \partial^\beta\partial_\tau G(\B.y) 
      \Big[K^{\mu\nu}(\B.y)\partial^{(y)}_\nu\partial^{(y)}_\mu 
         C^{\alpha\tau}(\B.r-\B.y)\Big] \ .
\nonumber
\end{eqnarray}

%
%

\section{Simplification of the ``nonlocal'' term}
\label{sec:simple}

For $\xi \ne 1$, the nontrivial integral on the LHS of \Eq{eq:Tab1} can be
further simplified. To see that, let us denote it by $I_{nt}$ and rewrite
it (omitting the $-1/2$ factor) with the explicit forms of the Kraichnan
operator and of the Green function
\begin{eqnarray}
\label{eq:non-local-1}
&& I_{nt} = -(\xi+2)\frac{D}{4\pi}\int\!\! d\B.y \, \frac{1}{y} 
     \partial^\beta\partial_\tau 
  \big[y^\xi\partial^2 C^{\alpha\tau}(\B.r-\B.y)\big] \\
&& \quad + \ \xi \frac{D}{4\pi}\int\!\! d\B.y \, \frac{1}{y} 
     \partial^\beta\partial_\tau 
  \big[y^{\xi-2}y^\mu y^\nu \partial_\mu\partial_\nu 
       C^{\alpha\tau}(\B.r-\B.y)\big] \ .
\end{eqnarray}
It is easy to verify that the tensors $\partial^2
C^{\alpha\tau}(\B.r-\B.y)$ and $y^\mu y^\nu\partial_\mu\partial_\nu
C^{\alpha\tau}(\B.r-\B.y)$ are divergence-free in both indices due to the
fact that $C^{\alpha\tau}(\B.r)$ itself is divergence free. Therefore, to
simplify the integrals in \Eq{eq:non-local-1}, we consider the generic
expression
\begin{equation}
  \int\!\! d\B.y \, \frac{1}{y} \partial^\beta\partial_\tau y^\lambda 
     X^{\alpha\tau}(\B.y) \ ,
\end{equation}
where $X^{\alpha\tau}(\B.y)$ is some divergence-free tensor and $\lambda$
is an arbitrary exponent. We also assume that $\lambda$ is such that the
integral is convergent and integration by parts is allowed. Then we may
write
\begin{eqnarray}
\label{eq:simp1}
&& \int\!\! d\B.y \,
  \frac{1}{y}\partial^\beta\partial_\tau 
    y^\lambda X^{\alpha\tau}(\B.y) \\
&&\quad  = \int\!\! d\B.y \, 
  \Big[ \lambda(\lambda-2)y^{\lambda-5}y^\beta y_\tau 
     X^{\alpha\tau}(\B.y) \nonumber \\
&&\quad +\ \lambda y^{\lambda-3}X^{\alpha\beta}(\B.y) 
        + \lambda y^{\lambda-3}y_\tau\partial^\beta 
   X^{\alpha\tau}(\B.y)\Big] \ .
\nonumber
\end{eqnarray}
The last formula can be simplified by using identity
\begin{eqnarray}
  0 &=& \int\!\! d\B.y \, \partial^\beta y^{\lambda-3}y_\tau
     X^{\alpha\tau}(\B.y) \\
  &=& (\lambda-3)\int\!\! d\B.y \, y^{\lambda-5}y^\beta y_\tau
     X^{\alpha\tau}(\B.y) \nonumber \\
  && + \int\!\! d\B.y \, y^{\lambda-3} X^{\alpha\beta}(\B.y)
     + \int\!\! d\B.y \, y^{\lambda-3}
     y_\tau\partial^\beta X^{\alpha\tau}(\B.y) \ , \nonumber
\end{eqnarray}
from which we get that for $\lambda \ne 3$
\begin{eqnarray}
\label{eq:simp2}
&& \int\!\! d\B.y \, \frac{1}{y}\partial^\beta\partial_\tau y^\lambda 
     X^{\alpha\tau}(\B.y) \\
&&\quad = -\frac{\lambda}{\lambda-3}\int\!\! d\B.y \, \big[
   y^{\lambda-3}X^{\alpha\beta}(\B.y) 
  + y^{\lambda-3}y_\tau\partial^\beta X^{\alpha\tau}(\B.y)\big] \ .
\nonumber
\end{eqnarray}
For $\lambda\ne 1$, we can even do better using the identity
\begin{eqnarray}
 0 &=& \int\!\!d\B.y \, \partial_\tau y^{\lambda-1}\partial^\beta 
   X^{\alpha\tau}(\B.y) \nonumber \\
   &=& (\lambda-1)\int\!\!d\B.y \, y^{\lambda-3}y_\tau\partial^\beta 
     X^{\alpha\tau}(\B.y) \ ,
\end{eqnarray}
which finally brings us to
\begin{equation}
\label{eq:simp3}
  \int\!\! d\B.y \, \frac{1}{y}\partial^\beta\partial_\tau y^\lambda 
     X^{\alpha\tau}(\B.y)
 = -\frac{\lambda}{\lambda-3}\int\!\! d\B.y \, y^{\lambda-3}
   X^{\alpha\beta}(\B.y) \ .
\end{equation}

Let us now apply \Eq{eq:simp3} to the integrals in \Eq{eq:non-local-1}.
Assuming that $\xi \ne 1$, we get
\begin{eqnarray}
\label{eq:non-local-2}
 I_{nt} &=& \frac{D\xi(\xi+2)}{4\pi(\xi-3)}\int\!\! d\B.y \, y^{\xi-3} 
     \partial^2 C^{\alpha\beta}(\B.r-\B.y) \\
 && - \frac{D\xi(\xi-2)}{4\pi(\xi-5)}\int\!\! d\B.y \, y^{\xi-5}
   y^\mu y^\nu \partial_\mu\partial_\nu C^{\alpha\beta}(\B.r-\B.y) \ .
\nonumber
\end{eqnarray}
To continue, we wish to turn the second integral into the same form of the
first integral. To accomplish that, consider the following identity:
\begin{eqnarray}
  0 &=& \int\!\!d\B.y \, \partial^\mu
 \big[ y^{\xi-3}y^\nu\partial_\mu\partial_\nu 
      C^{\alpha\beta}(\B.r-\B.y)\big] \\
 &=&(\xi-3)\int\!\!d\B.y \, y^{\xi-5}y^\mu y^\nu\partial_\mu\partial_\nu
      C^{\alpha\beta}(\B.r-\B.y) \nonumber \\
  && + \int\!\! d\B.y \, y^{\xi-3}\partial^2 
       C^{\alpha\beta}(\B.r-\B.y) \nonumber \\
  && + \int\!\! d\B.y \, y^{\xi-3}y^\nu \partial_\nu 
       \partial^2 C^{\alpha\beta}(\B.r-\B.y) \ ,
\nonumber
\end{eqnarray}
which gives us
\begin{eqnarray}
&& \int\!\! d\B.y \, y^{\xi-5}
   y^\mu y^\nu \partial_\mu\partial_\nu C^{\alpha\tau}(\B.r-\B.y) \\
&& \quad = -\frac{1}{\xi-3}\int\!\! d\B.y \, \big[ 
   y^{\xi-3}\partial^2 C^{\alpha\beta}(\B.r-\B.y) \nonumber \\
&& \qquad +\ y^{\xi-3}y^\nu\partial_\nu \partial^2 
   C^{\alpha\beta}(\B.r-\B.y)\big] \ .
\nonumber
\end{eqnarray}
Additionally, we have
\begin{eqnarray}
 0 &=& \int\!\!d\B.y \, \partial_\nu y^{\xi-3}y^\nu \partial^2
    C^{\alpha\beta}(\B.y) \\
   &=& \xi \int\!\!d\B.y \, y^{\xi-3} \partial^2 C^{\alpha\beta}(\B.y)
 + \int\!\!d\B.y \, y^{\xi-3}y^\nu \partial_\nu \partial^2 
    C^{\alpha\beta}(\B.y) \ ,
\nonumber 
\end{eqnarray}
and so finally we obtain
\begin{eqnarray}
&& \int\!\! d\B.y \, y^{\xi-5}
   y^\mu y^\nu \partial_\mu\partial_\nu C^{\alpha\tau}(\B.r-\B.y) \\
&&\quad = \frac{\xi-1}{\xi-3}\int\!\! d\B.y \, y^{\xi-3}
     \partial^2 C^{\alpha\beta}(\B.r-\B.y) \ .
\nonumber
\end{eqnarray}
Substituting this into \Eq{eq:non-local-2}, we arrive at the final result
for $I_{nt}$:
\begin{equation}
 I_{nt}  = \frac{12\xi D}{(\xi-3)(\xi-5)}
      \int\!\! d\B.y \, G(\B.y) y^{\xi-2} 
      \partial^2 C^{\alpha\beta}(\B.r-\B.y) \ .
\end{equation}

%
%

\section{Calculation of $A(\lambda; \ell, \xi)$}
\label{sec:A}

The prefactor $A(\lambda; \ell, \xi)$ was defined by
\begin{eqnarray}
&& A(\lambda; \ell, \xi) r^{\lambda+\xi} Y_{\ell m}(\hat{\B.r}) \\
&& \quad \equiv \ \int\!\! d\B.y \, G(\B.r-\B.y) 
          |\B.r-\B.y|^{\xi-2} y^\lambda Y_{\ell m}(\hat{\B.y}) \nonumber \ .
\end{eqnarray}
Due to the isotropy of the integral, it is $m$ independent and therefore we
specialize to the $m=0$ where the $Y_{\ell m}(\hat{\B.y})$ is proportional
to the Legendre polynomial $P_\ell(\hat{\B.y}\cdot\hat{\B.z})$. Setting
$\B.r=\hat{\B.z}$, the unit vector in the $z$ direction, we write the
integral in spherical variables $(y,\theta, \phi)$ and perform the trivial
$\phi$ integration (for $\B.r = \hat{\B.z}$, the integrand is independent
of $\phi$). We arrive at
\begin{eqnarray}
 A(\lambda; \ell, \xi) &=& -\frac{1}{2}\int_0^\infty \!\! dy\, y^{2+\lambda}
    \int_{-1}^1 \!\! d(\cos\theta) \\
 && \times\ (y^2 - 2y\cos\theta + 1)^{\frac{\xi-3}{2}} 
   P_\ell(\cos\theta) \ . \nonumber
\end{eqnarray}

Using the standrad tricks of Feynmann integrals, one can express this
integral (at least in the $\ell=0$ case) in terms of gamma functions. Here,
however, we choose to calculate this integral directly by a strightforward
expansion of the integrand. This procedure underlines the connection
between the pole structure and the anisotropy label $\ell$.

Let us therefore turn the annoying $(y^2-2y\cos\theta +1)^{(\xi-3)/2}$ term
into a Taylor series in $2y\cos\theta/(1+y^2)$,
\begin{eqnarray}
&&  A(\lambda; \ell, \xi) = -\frac{1}{2}
     \sum_{n=0}^\infty \frac{(-1)^n 2^n}{n!} \nonumber \\
&&\quad \times \ \left( \frac{\xi-3}{2}\right)\cdot
     \left( \frac{\xi-3}{2} - 1 \right) \cdots \cdot
     \left( \frac{\xi-3}{2} - n + 1 \right) \nonumber \\
&&\quad \times \ B(n,\ell) C(\lambda, n, \xi)  \ , 
\label{eq:A-mid}
\end{eqnarray}
where $B(\lambda, n, \xi)$ and $C(n, \ell)$ are two one-dimensional
integrals that are given by
\begin{eqnarray}
\label{def:B-int}
  B(n,\ell) &\equiv& \int_{-1}^1 \!\! d(\cos\theta) \, \cos^n\!\theta \, 
     P_\ell(\cos\theta) \ , \\
  C(\lambda, n, \xi) &\equiv& \int_0^\infty \!\! dy\, 
     y^{2+n+\lambda} (1+y^2)^{\frac{\xi-3}{2}-n} \ .
\label{def:C-int}
\end{eqnarray}
Before calculating $B(n, \ell)$, we notice that it vanishes for $n<\ell$
and for $n$s that are of different parity than $\ell$. The first
observation is a simple manifestation of the orthogonality of the Legendre
polynomials intimately connected to the orthogonality of different
irreducible representations of the SO($3$) group. For $n>\ell$ with the
same parity, we can use the well-known identity
\begin{equation}
  P_\ell(x) = \frac{1}{2^\ell \ell!}\frac{d^\ell}{dx^\ell}
     (x^2 - 1)^\ell \ ,
\end{equation}
from which we get, after simple integration by parts, that $B(n,\ell)$ is
given by
\begin{eqnarray}
  B(n,\ell) &=& (-1)^\ell 
   \frac{n(n-1)\cdots(n-\ell+1)}{2^\ell \ell!} \\
  &&\times\ \int_{-1}^{1}\!\! dx\, x^{n-\ell}(x^2 - 1)^\ell dx \ .
\nonumber
\end{eqnarray}
The above integral can be done explicitly leading us to the final result
\begin{eqnarray}
  B(n,\ell) &=& \frac{n(n-1)\cdots(n-\ell+1)}{2^\ell \ell!} \\
   && \times \ \sum_{k=0}^n {\ell \choose k} \frac{2 (-1)^k}{n+2k-\ell} \ .
\nonumber
\end{eqnarray}
  
We now calculate the second integral by dividing the integration regime
$[0,+\infty]$ in \Eq{def:C-int} into a $0<y<1$ part and a $1<y<\infty$
part. In each part we expand the $(1+y^2)^{(\xi-3)/2-n}$ term in $y$
and $1/y$, respectively, and perform the integration. After adding up these
terms again, we arrive at the following sum
\begin{eqnarray}
&& C(\lambda, n, \xi) = \sum_{k=0}^\infty \frac{1}{k!}
    \left( \frac{\xi-3}{2} - n \right) \\
&&\quad \times\ \left( \frac{\xi-3}{2} - n -1 \right) \cdot \ldots \cdot
    \left( \frac{\xi-3}{2} - n - k + 1\right) \nonumber \\ 
&& \quad\times\ \left[ \frac{1}{\lambda+3+n+2k} - 
     \frac{1}{\lambda+\xi-n-2k} \right] \ .
\nonumber
\end{eqnarray}
Plugging these results back in \Eq{eq:A-mid}, we get
\begin{eqnarray}
\label{eq:A-mid2}
&&  A(\lambda; \ell, \xi) = -\frac{(-1)^\ell}{2}
     \sum_{n=\ell,\ell+2,\ldots}^\infty 
     \sum_{k=0}^\infty \frac{2^n}{n!k!} \\
&&\quad \times \ \left( \frac{\xi-3}{2}\right)\cdots
     \left( \frac{\xi-3}{2} - n -k + 1 \right) \nonumber \\
&& \quad\times\left[ \frac{1}{\lambda+3+n+2k} - 
     \frac{1}{\lambda+\xi-n-2k} \right] B(n,\ell) \ .
\nonumber
\end{eqnarray}
To expose the structure of the poles in the above formula, it is convenient
to change the order of summation by defining the new indices $q,j$,
\begin{eqnarray}
\label{def:qj}
  q &\equiv& \frac{n-\ell}{2} + k = 0,1,2,\ldots \ , \\
  j &\equiv& \frac{n-\ell}{2}     = 0,1,2,\ldots,q \ , \nonumber
\end{eqnarray}
and obtain
\begin{eqnarray}
\label{eq:A-final}
&& A(\lambda; \ell, \xi)  = -\frac{(-1)^\ell}{2}\sum_{q=0}^\infty 
   a_q(\ell,\xi) \\
&&\quad \times \ \left[ \frac{1}{\lambda+3+\ell+2q} - 
     \frac{1}{\lambda+\xi-\ell-2q} \right] \ , \nonumber
\end{eqnarray}
where $a_q(\ell,\xi)$ are given by
\begin{eqnarray}
\label{def:aq}
&& a_q(\ell, \xi) \equiv \sum_{j=0}^q 
   \frac{2^{\ell+2j}}{(\ell+2j)!(q-j)!} \\
&& \quad \times\ \Big( \frac{\xi-3}{2}\Big)\cdot\ldots \cdot
     \Big( \frac{\xi-3}{2} - \ell-j - q + 1 \Big) B(\ell + 2j, \ell) 
\nonumber \ .
\end{eqnarray}

Equation \ref{eq:A-final} shows that $A(\lambda; \ell, \xi)$ is given as an
infinite series of poles in $\lambda$. For $\xi>0$ this series can be shown
to converge although for small values of $\xi$ the convergence is very
slow.  In the special case of $\xi=2$, the series is truncated after the
first pole. To see why this is so, return to the original definition of
$A(\lambda; \ell, \xi)$ and set $\xi=2$:
\begin{equation}
  A(\lambda; \ell, 2)r^{\lambda+2}Y_{\ell m}(\hat{\B.r}) =
    \int\!\!d\B.y \,G(\B.r-\B.y) y^\lambda Y_{\ell m}(\hat{\B.y}) \ .
\end{equation}
But since
\begin{equation}
  y^\lambda Y_{\ell m}(\hat{\B.y}) = 
   \frac{1}{(\lambda+2-\ell)(\lambda+3+\ell)} 
   \partial^2 y^{\lambda+2} Y_{\ell m}(\hat{\B.y}) \ ,
\end{equation}
then from the definition of the Green function we get
\begin{equation}
 A(\lambda; \ell, 2) = \frac{1}{(\lambda+2-\ell)(\lambda+3+\ell)} \ .
\end{equation}

%
%

\end{multicols}

\end{document}